\documentclass[journal]{IEEEtai}

\usepackage[colorlinks,urlcolor=blue,linkcolor=blue,citecolor=blue]{hyperref}

\usepackage{color,array}

\usepackage{graphicx}
\usepackage{times}
\usepackage{latexsym}
\usepackage{amssymb}
\usepackage{mdwlist}
\usepackage{float}
\usepackage{booktabs}
\usepackage{amsmath,amssymb,mathrsfs}
\usepackage{bbding}
\usepackage{makecell}
\usepackage{longtable}
\usepackage{rotating}
\usepackage{multirow}
\usepackage{array}
\usepackage{color}
\usepackage{bm}
\usepackage{hyperref}
\usepackage{lscape}
\usepackage{pdflscape}
\usepackage{stfloats}
\usepackage{subfig}
\usepackage{caption}
\usepackage{colortbl}
\usepackage{xcolor}
\usepackage{algorithm}
\usepackage{algorithmic}
\usepackage{tabularx}

\begin{document}

\title{Hy-DeFake: Hypergraph Neural Networks for Detecting Fake News in Online Social Networks} 

\author{Xing Su, Jian Yang, Jia Wu, \IEEEmembership{Senior Member, IEEE,} Zitai Qiu
\thanks{X. Su, J. Yang, J. Wu, and Z. Qiu are with the School of Computing, Macquarie University, Sydney, Australia. 
E-mail: \{xing.su2, zitai.qiu\}@students.mq.edu.au, \{jian.yang, jia.wu\}@mq.edu.au.}
\thanks{This paragraph will include the Associate Editor who handled your paper.}}

\markboth{Journal of IEEE Transactions on Artificial Intelligence, Vol. 00, No. 0, Month 2020}
{First A. Author \MakeLowercase{\textit{et al.}}: Bare Demo of IEEEtai.cls for IEEE Journals of IEEE Transactions on Artificial Intelligence}

\maketitle

\begin{abstract}
Nowadays social media is the primary platform for people to obtain news and share information. Combating online fake news has become an urgent task to reduce the damage it causes to society. Existing methods typically improve their fake news detection performances by utilizing textual auxiliary information (such as relevant retweets and comments) or simple structural information (\textit{i.e.}, graph construction). However, these methods face two challenges. First, an increasing number of users tend to directly forward the source news without adding comments, resulting in a lack of textual auxiliary information. Second, simple graphs are unable to extract complex relations beyond pairwise association in a social context. Given that real-world social networks are intricate and involve high-order relations, we argue that exploring beyond pairwise relations between news and users is crucial for fake news detection. Therefore, we propose constructing an attributed hypergraph to represent non-textual and high-order relations for user participation in news spreading. We also introduce a hypergraph neural network-based method called Hy-DeFake to tackle the challenges. Our proposed method captures semantic information from news content,  credibility information from involved users, and high-order correlations between news and users to learn distinctive embeddings for fake news detection. The superiority of Hy-DeFake is demonstrated through experiments conducted on four widely-used datasets, and it is compared against eight baselines using four evaluation metrics. 
\end{abstract}

\begin{IEEEImpStatement}
Online fake news detection algorithms play a crucial role in combating the spread of misinformation. They offer automatic and timely news identification by swiftly analyzing vast amounts of data, thereby reducing human workload. However, existing efforts have not fully leveraged the complicated relations of news and users within social networks, which limits the precision of fake news detection. The hypergraph neural network based method we introduced in this work overcame such limitations. With accuracy rates exceeding 92\% on all tested datasets, our method surpasses all the baseline algorithms. This enhancement not only fortifies the reliability of online information but also empowers platforms, fact-checkers, and users in combating the pervasive threat of fake news. 
\end{IEEEImpStatement}


\section{Introduction}

\IEEEPARstart{N}{owadays,} social media platforms such as Facebook and Twitter have emerged as primary sources for accessing timely news and sharing information. While these platforms offer the advantage of instant access to real-time news from diverse perspectives, the ease of engagement and interaction has led to the rapid proliferation of fake news. This spread of misinformation has resulted in significant physical and mental harm to individuals. For instance, over 5,800 people were admitted to hospital after falling victim to a piece of COVID-19 fake news, which claims that alcohol with high concentration can sanitize the body and eradicate the virus \cite{islam2020covid}. Fake news not only poses a threat to public health but also undermines trust in governments and alters public perception \cite{10.1145/3055153}. For example, research has shown that fake news circulating on social media played a role in shaping the vote for the Brexit referendum \cite{bastos2019brexit}. During the Brexit campaign, a significant amount of false information circulated on social platforms, with pro-Leave fake news spreading more widely on Twitter than pro-Remain fake news. The widespread dissemination of fake news has had detrimental social consequences. Therefore, it is an imminent task to identify fake news online to ensure the public receives trustworthy information and reliable guidance. 
  
\begin{figure}[!tbp]
\centering
\includegraphics[scale=0.25]{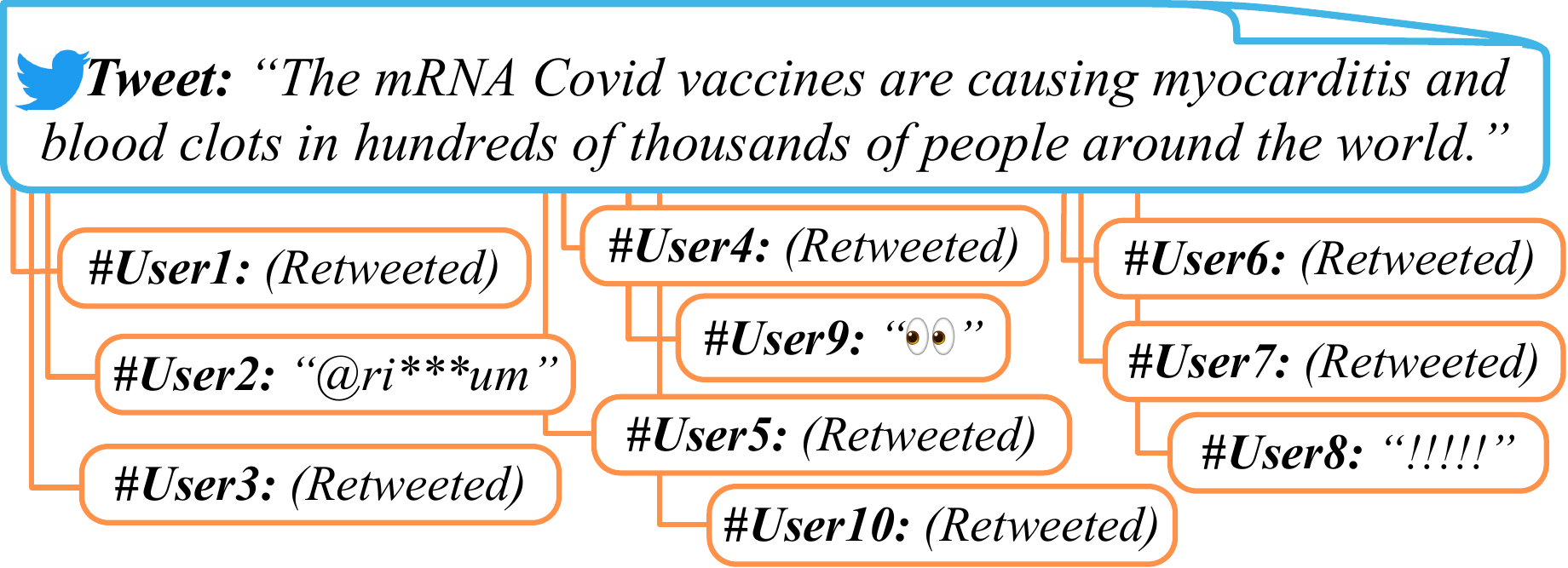} 
\caption{An illustrative example of a piece of news spreading by retweeting. }
\label{fig:toy}
\end{figure}

Various methods have been developed for the detection of fake news in online social networks \cite{shu2017fake,guo2020future}. A simple technique to identify fake news is through text classification, but the results indicate that these approaches have limited performance since they only consider news content itself and overlook the social context \cite{perez-rosas-etal-2018-automatic}. To achieve more accurate results, some studies adopt textual data from social context as auxiliary information, such as comments, tweets, and retweets of source news \cite{defend19shu,rao-etal-2021-stanker,ma-etal-2018-rumor}. However, these methods face a practical problem: many users tend to retweet the news directly without adding comments or expressing their stances, which results in the absence of textual side information. Let's take the spreading of a piece of COVID-19 news as an example, as shown in Fig. \ref{fig:toy}. Most users participate in sharing this news by directly retweeting it, such as \textit{\#User1}, \textit{\#User7}. The comments published during retweeting are often insignificant, like the symbols from \textit{\#User8} and the expression from \textit{\#User9}. As of May 11, 2023, the news in Fig. \ref{fig:toy} has 10.2K retweets without comments, but only 380 retweets with comments. Therefore, the challenge in real-world fake news detection is how to exploit non-textual auxiliary information from social context for accurate results, which we refer to as \textit{challenge \#1}. 

To address the \textit{challenge \#1}, some models utilizing non-textual structural information have been proposed. For example, some models construct propagation graphs for tweets and retweets \cite{bian2020rumor}, while others employ heterogeneous graphs for news and users \cite{huang2020heterogeneous}. These approaches have demonstrated notable performance in identifying fake news, as the extracted structural information aligns well with the network formed by news spreading or news-user interaction in social media. However, these methods still have a limitation: they typically capture pairwise relations in social networks. In real-world social networks, there are numerous high-order and intricate relations that these methods fail to capture beyond pairwise relations. For instance, a pairwise relation can represent the relationship between a user and a single piece of news he or she posts, but it cannot represent the relations among multiple news posts by the same user. Thus, extracting high-order relations from social networks to gather in-depth information for enhancing fake news detection presents \textit{challenge \#2}. 

\begin{figure}[!bh]
    \centering
    \includegraphics[scale=0.15]{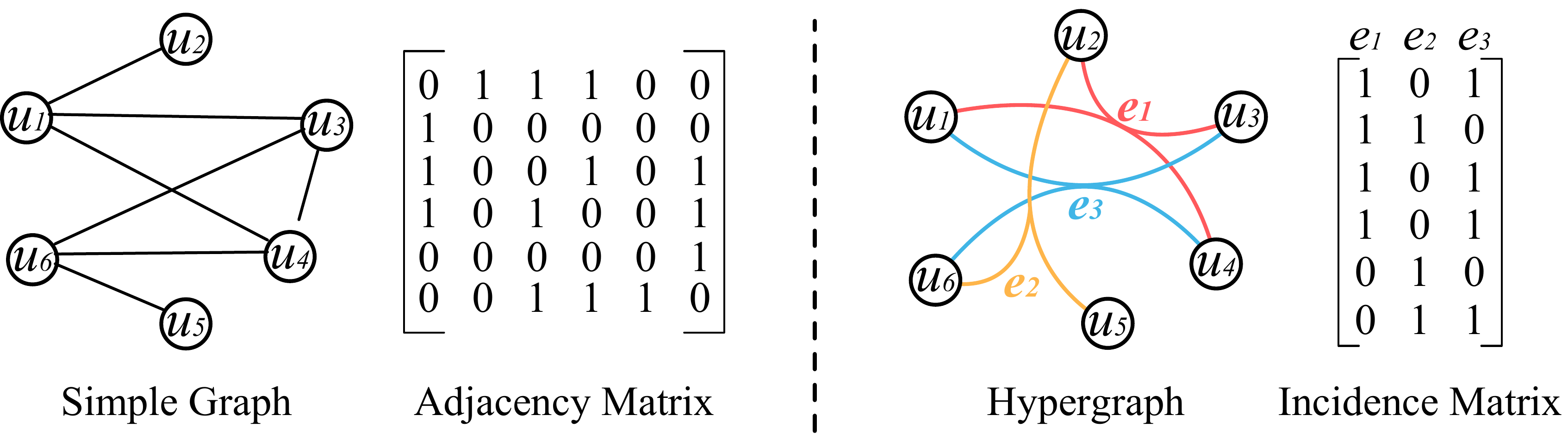}
    \caption{Difference between simple graph and hypergraph.}
    \label{fig:graphcompare}
\end{figure}

In this work, to address the \textit{challenge \#1}, we leverage user information to complement the plain text of news content. Given the participatory and user-driven nature of news dissemination on online social networks, we argue that users' engagement and dissemination patterns may vary between fake and real news, providing valuable insights for fake news detection. Specifically, to incorporate non-textual information from the social context, we aim to extract the inherent attributes of users as well as the structures of both news and users during news spreading. To tackle the \textit{challenge \#2}, we employ the concept of hypergraph to capture intricate high-order relations between news and users \cite{bretto2013hypergraph}. Hypergraphs are generalizations of simple graphs that can connect an arbitrary number of nodes in an edge. As depicted in Figure \ref{fig:graphcompare}, a simple graph only connects two nodes in an edge and its adjacency matrix illustrates the pairwise relation between nodes. However, the incidence matrix of a hypergraph demonstrates that multiple nodes are assigned to one edge, enabling the extraction of high-order information. Therefore, in this work users are abstracted as nodes, and news is represented as hyperedges. This hypergraph construction allows us to capture the intricate high-order relation between a source news piece and the users who participated in its dissemination within each hyperedge, enabling the learning of distinctive information for fake news detection. 
Within this hypergraph, node attributes represent user properties related to credibility, while hyperedge attributes encompass the textual contents of news. 

To comprehensively learn the latent high-order correlation between news and users in our constructed hypergraph and overcome the two challenges, we propose a method based on \textbf{\underline{Hy}}pergraph neural networks for \textbf{\underline{De}}tecting \textbf{\underline{Fake}} news, abbreviated as Hy-DeFake. Hy-DeFake consists of four main parts, utilizing the attributed hypergraph as input: (1) News semantic channel: it updates hyperedge features by learning semantic embeddings of news contents, as each hyperedge represents a piece of news; (2) User credibility channel: it learns node features through embedding both user credibility information and the high-order structural information between news and users; (3) Mutual information-based feature fusion: it incorporates the semantic embeddings of news (\textit{i.e.}, hyperedges) and the credibility embeddings of relevant users (\textit{i.e.}, nodes) based on maximizing their mutual information in high-dimensional space; (4) Fake news detection: it models our task as hyperedge classification and classifies the news based on the integrated embeddings. Specifically, Hy-DeFake simultaneously trains the news semantic channel and the user credibility channel to acquire diverse information for identifying fake news. The news semantic channel leverages a language model to exploit semantic embeddings of news. The user credibility channel employs a hypergraph autoencoder, where the encoder is a hypergraph convolution network, that learns the high-order information between nodes and hyperedges. Through this process, the two channels learn the semantic embeddings of news and user embeddings, which capture both user credibility and high-order correlation between news and users, respectively. This work presents the following key contributions:
\begin{itemize}
\item We introduce an approach by constructing an attributed hypergraph to represent the process of news spreading in online social networks. By abstracting fake news detection as hyperedge classification, we capture the intricate high-order relation between news and users in social contexts, enabling us to achieve accurate results. 
\item The proposed Hy-DeFake utilizes hypergraph neural networks for fake news detection. It effectively captures the credibility information of users and the high-order correlation between news and users. Both of these aspects provide distinctive information that contributes to fake news detection. 
\item Extensive experiments demonstrate that Hy-DeFake generally surpasses eight baseline methods on four real-world datasets from different domains. 
\item We uncover a positive correlation between news authority and user credibility. Users who spread fake news exhibit more intensive interaction compared to those who spread real news, resulting in the formation of a denser community. 
\end{itemize}

This paper consists of the following parts. Section \ref{sec:relatedwork} reviews the relevant literature on hypergraph learning, fake news detection, and the distinctions between Hy-DeFake and existing methods. The relevant definitions and hypergraph construction are formulated in Section \ref{sec:formulation}. Following that, section \ref{sec:framework} presents each part of the proposed framework, and section \ref{sec:experiments} shows experimental evaluations. Finally, we draw conclusions in this work and outline our future work in Section \ref{sec:conclusion}.

\section{Related work}
\label{sec:relatedwork}
To give an overview of the relevant research, this section first reviews the work on hypergraph learning and its applications, then further discusses the work on fake news detection.  

\subsection{Hypergraph Learning}
The development of graph neural networks (GNN) \cite{9046288} has demonstrated its success in various tasks and has garnered significant attention for its ability to uncover structural relations \cite{10.1145/3535101,su2022comprehensive}. However, traditional graph learning methods primarily focus on pairwise connections, limiting their capacity to express complex high-order relations that extend beyond pairwise associations. Therefore, as a promising and flexible technique for modeling complex high-order relations, hypergraph learning is emerging recently. The first work that applies GNNs to hypergraphs is HGNN \cite{feng2019hypergraph}. To leverage complex and high-order relations for better representation learning, it designs a hyperedge convolution operation. HyperGCN \cite{Yadati2019HyperGCN} trains a GCN on hypergraph by adopting tools from the spectral theory of hypergraphs and extending its faster variant. Hyper-SAGNN \cite{zhang2019hyper} develops a self-attention based GNN for homogeneous and heterogeneous hypergraphs. DHGNN (Dynamic Hypergraph Neural Networks) \cite{jiang2019dynamic} is a framework that exploits adjusted feature embeddings to dynamically update hypergraph structure. HyperGCL \cite{wei2022_hypergcl} is a hypergraph generative model that applies contrastive learning for robust and fair hypergraph representation learning. 

Due to its capacity to capture complicated high-order relations, hypergraph neural networks have found applications in various domains. For instance, SHINE \cite{Luo2022shine} is a subhypergraph inductive neural network for simultaneously capturing functional relations for genes and pathways. HHGR \cite{zhang2021double} and DH-GCN \cite{han2022dh} are two examples of exploiting hypergraph neural networks for the social recommendation. GroupNet \cite{Xu_2022_CVPR} introduces a multiscale hypergraph neural network for trajectory prediction. We can conclude that a hypergraph neural network is suitable for numerous real-world scenarios, particularly in the context of complex social networks. However, its application in the domain of detecting fake news in online social networks is relatively limited. Available approaches utilizing hypergraphs for fake news detection are quite limited \cite{jeong2022nothing,ding2020more}. HyperGAT \cite{ding2020more} primarily focuses on news content, lacking the integration of high-order relations of both users and news. HGFND \cite{jeong2022nothing} requires numerous types of hyperedges, increasing the difficulty of dataset construction and impacting real-world practicality. Moreover, it still relies on vanilla GNNs to learn hypergraph representations and is limited in capturing latent high-order relations. Thus, this work is proposed to address such issues.

\subsection{Fake News Detection}
\subsubsection{Text-based Detection}
From the perspective of news contents, text-based approaches detect fake news by distinguishing linguistic features and writing styles between real and fake news. P{\'e}rez-Rosas \textit{et al.}, for instance, \cite{perez-rosas-etal-2018-automatic} introduced a text-based method focusing on learning linguistic features to identify fake news. In FakeBERT \cite{kaliyar2021fakebert}, Kaliyar \textit{et al.} integrated BERT \cite{devlin2018bert} with multiple parallel blocks of 1d-CNN for fake news detection. Apart from analyzing the textual contents of the news itself, auxiliary textual information is also integrated to enhance fake news detection. dEFEND \cite{defend19shu}, for example, adopts a hierarchical co-attention mechanism to capture the explainable sentences from both news and the relevant comments. Similarly, STANKER \cite{rao-etal-2021-stanker} utilizes auxiliary features through comments and leverages the level-grained attention-masked BERT model to uncover fake news. Furthermore, mining stance or emotion in news content can provide side information as well. For example, Zhang \textit{et al.} \cite{dualemo2021zhang} specifically examined the impact of emotional cues in real and fake news, and incorporated dual emotion features to represent both individual emotions and the relationship between them. Yang \textit{et al.} proposed two tree-structured models (TD-MIL and BU-MIL) \cite{10.1145/3477495.3531930} to simultaneously verify rumorous claims and detect the stances behind the relevant posts. Sheng \textit{et al.} \cite{sheng-etal-2022-zoom} consider both macro and micro news environments for fake news identification. FTT \cite{hu-etal-2023-learn} detects fake news by discovering topics in the news and forecasting temporal trends. While text-based models utilize textual information and auxiliary features for fake news detection, they commonly analyze each news piece in isolation, overlooking the intrinsic news-spreading mechanism. As a result, they may neglect valuable propagation information. These methods often lack user information which is crucial for understanding the social context to accurately identify fake news. Consequently, the performance of these approaches becomes limited. 

\subsubsection{Graph-based Detection}
Due to the compatibility between the propagation nature of news and the structural property of graphs, graph learning shows high-quality results in fake news detection. In graph-based methods, researchers typically extract structural relations of words or sentences, construct news propagation graphs or incorporate social contexts (such as user information) to detect fake news from both textual and structural information perspectives. For instance, Yao \textit{et al.} used GCN \cite{kipf2017semisupervised} to learn news-level structure information. They constructed graphs to represent individual news pieces \cite{yao2019graph}. Wu \textit{et al.} \cite{10.1145/3477495.3531850} developed a causal graph to depict the relations between news and evidence in a counterfactual inference framework. Additionally, Ma \textit{et al.} \cite{ma-etal-2018-rumor} created a tree-structured graph based on tweets and utilized recursive neural networks to represent the graph in both top-down and bottom-up fashion. Similarly, Bi-GCN \cite{bian2020rumor} is a bi-directional graph neural network that understands patterns of news propagation and dispersion by GCN \cite{kipf2017semisupervised}. To sum up, graph-based approaches have the ability to represent the textural and structural properties of debunking fake news. 

To fully harness the potential of graph structure for accurate results, researchers also incorporate structural side information. CompareNet \cite{hu2021compare} exploits knowledge graphs to detect fake news. For capturing various connections, Huang \textit{et al.} \cite{huang2020heterogeneous} and HGAT \cite{ren2021fake} are two works constructing heterogeneous information networks (HINs). For considering user information in social contexts, UPFD \cite{dou2021user} considers user preferences by summarizing historical posts, Us-DeFake \cite{su2023wsdm} captures news-user relations in a dual-layer graph, and PESTO \cite{min-ananiadou-2023-pesto} defines a post-user fusion network to model the rumor patterns from both post diffusion and user social networks. MFAN \cite{ijcai2022p335} integrates texts, visions, and social graph features in a unified framework, to obtain complementary relations between different modalities. Furthermore, Mehta \textit{et al.} \cite{mehta-etal-2022-tackling} used inference operators to reveal interactions like the similarity between news content and user engagement patterns. SureFact \cite{10.1145/3534678.3539277} is based on multiple heterogeneous subgraphs that extract diverse information from claims, posts, keywords, and users. Similarly, Jin \textit{et al.} constructs a claim-evidence graph by news articles, posts, and users, and then proposes \cite{jin2022towards} a fine-grained reasoning model FinerFact. PSIN \cite{10.1145/3485447.3512163} models the post and user interactions through a divide-and-conquer strategy to detect fake news. DECOR \cite{wu2023decor} refines social graphs by degree-corrected stochastic blockmodels for accurate fake news detection. 

Albeit the variety of methods considers user information in social contexts, they are inadequate in modeling the complex high-order relations between news and users. Therefore, we propose Hy-DeFake to address the aforementioned problems. Hy-DeFake goes beyond considering only textual information and incorporates structural information within a social context. Specifically, it also captures the high-order correlation between news and users to acquire distinctive representations for accurate identification of fake news.

\section{Preliminaries} 
\label{sec:formulation}
In this section, we begin by describing the notions and definitions utilized in this paper, and then present how a social network with news and users is modeled as a hypergraph. Finally, we state our problem of detecting fake news. 

\subsection{Notations and Definition}
Let $\mathcal{N} = \{n_{1}, n_{2}, \cdots, n_{t}\}$ denote the news set, where $t$ represents the number of news and $n_{i}$ represents the textual contents of news $i$. Let $\mathcal{U} = \{u_{1}, u_{2}, \cdots, u_{m}\}$, where $m$ is the number of users. The corresponding user attributes are defined as $\bm{X}\subseteq\mathbb{R}^{m\times d}$, where $d$ refers to the dimension of user attribute vectors. 

\textit{Hypergraph Definition}: Hypergraph is the generalized graph, it can be formulated as $\mathcal{G} = (\mathcal{V}, \mathcal{E}, \bm{X})$, where $\mathcal{V}$ and $\bm{X}$ indicate the node set and the corresponding node attributes. $\mathcal{E}$ signifies the set of hyperedges, where each hyperedge connects to a subset of nodes rather than two nodes. The diagonal matrices of edge degrees and node degrees can be represented as $\bm{D}_{e}$, $\bm{D}_{v}$, respectively. $\bm{H} = |\mathcal{V}| \times |\mathcal{E}|$ denotes incidence matrix, where $\bm{h}(v,e) = 1$ if the node $v$ is on hyperedge $e$ (\textit{i.e.}, $v\in e$), else $\bm{h}(v,e) = 0$.

\subsection{Hypergraph Construction}
\begin{figure}[!htbp]
\centering
\includegraphics[scale=0.22]{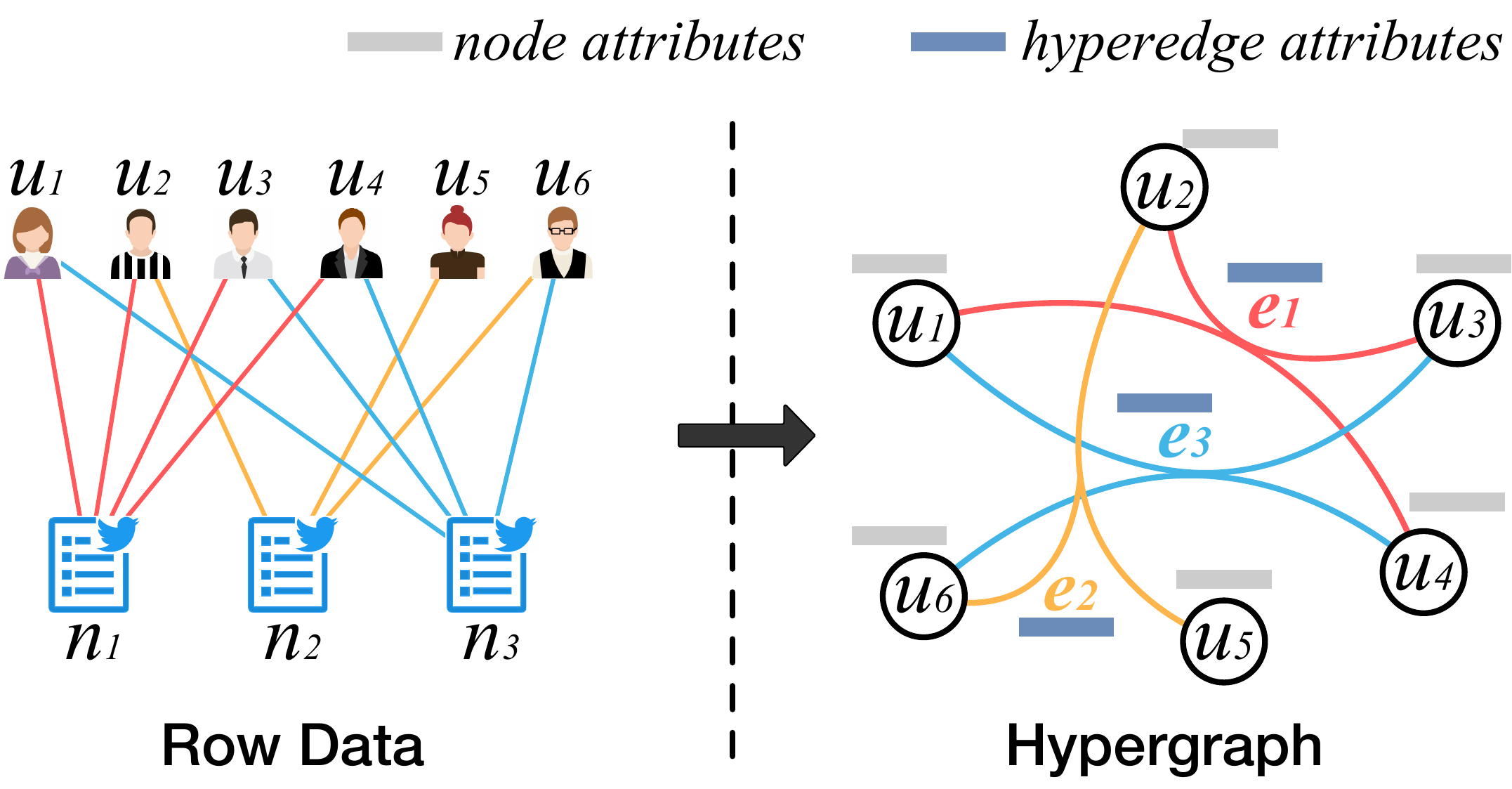} 
\caption{Hypergraph construction. }
\label{fig:construction}
\end{figure}

In this work, to model the relation between each news piece and the users involved in its spreading in online social networks for the purpose of debunking fake news, we define an attributed hypergraph as $\mathcal{G} = (\mathcal{U}, \mathcal{E}, \bm{X}, \mathcal{N})$ to represent the beyond pairwise relations. Here the node set $\mathcal{U}$ denotes the users who are involved in news spreading. The hyperedge set $\mathcal{E}$ denotes the relation between news and users, where each hyperedge $e_{i}$ represents a piece of news $i$ and can connect more than two users who are involved in the spread of this news. $\bm{X}$ and $\mathcal{N}$ represents attributes of nodes (\textit{i.e.}, users) and hyperedges (\textit{i.e.}, news), respectively. The node attributes of users are user properties, such as the number of followers and verified status, while the hyperedge attributes are the textual contents of news. Figure \ref{fig:construction} shows the changes in row data after hypergraph construction. 

Thus, the task of this work is abstracted as distinguishing news in a hypergraph, \textit{i.e.}, hyperedge classification task. Treating fake news detection as a binary classification task, we associate a binary label $y\in\{0, 1\}$ to each hyperedge $e_{i}$ in a hypergraph. A label of 0 is assigned to indicate that the corresponding node represents real news, whereas a label of 1 is assigned to indicate fake news. Thus, the hypergraph for fake news detection can be denoted as $\mathcal{G} = (\mathcal{U}, \mathcal{E}, \bm{X}, \mathcal{N}, \mathcal{Y})$, where $\mathcal{Y}$ represents the label set assigned to the news.

\section{Methodology}
\label{sec:framework}
\begin{figure*}[!tbp]
\centering
\includegraphics[scale=0.17]{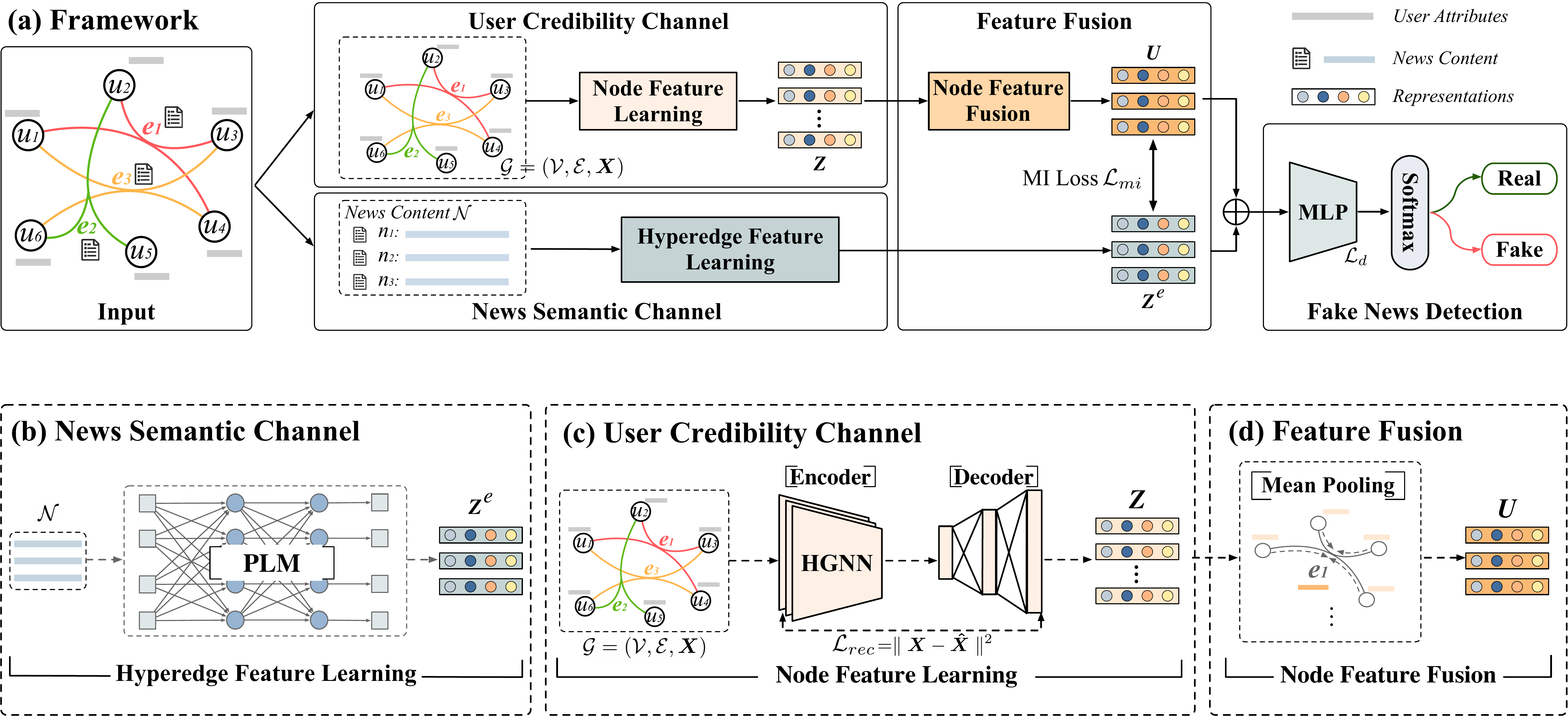} 
\caption{\textbf{The Framework of Hy-DeFake}. \textbf{(a) Framework} shows the overall architecture of Hy-DeFake. It takes a hypergraph of news and users as input. Firstly, Hy-DeFake designs two channels to learn features from news and users, respectively. Hyperedge attributes (\textit{i.e.}, news content) are fed into the news semantic channel to learn semantic features of textual data. Hypergraph with node attributes is fed into the user credibility channel to learn node features with credibility information of users and high-order structural information between news and users. After that, a mutual information-based feature fusion module fuses hyperedge features and representative node features to obtain integrated embeddings for fake news detection. Finally, the fused embeddings are fed into the fake news detection component to classify the news into real and fake. \textbf{(b) News Semantic Channel} shows the learning process of hyperedge attributes $\mathcal{N}$ to update news features $\bm{Z}^{e}$ in a pre-trained language model (PLM). \textbf{(c) User Credibility Channel} shows the details in this channel, it adopts a hypergraph neural network (HGNN) in Autoencoder architecture to learn the node features $\bm{Z}$. \textbf{(d) Feature Fusion} fuses $\bm{Z}$ on each hyperedge to aggregate the representative user embeddings $\bm{U}$ for further fusion with news embeddings. }
\label{fig:framework}
\end{figure*}

In this section, we provide a detailed explanation of our proposed method Hy-DeFake. The model takes our constructed hypergraph as input. To fully capture credibility information from users and semantic information from news, we design two channels to learn user and news information individually. As depicted in Fig. \ref{fig:framework}, Hy-DeFake comprises four parts: 1) the news semantic channel, 2) the user credibility channel, 3) the mutual information-based feature fusion, and 4) the fake news detection. The news semantic channel adopts a language model to update hyperedge features and comprehend the semantics of news. The user credibility channel exploits an unsupervised hypergraph autoencoder to acquire credibility features of users, preserving the high-order information between news and users. Then, the mutual information-based feature fusion module aggregates node features users involved in each piece of news to obtain representative credibility embeddings of users. This module also maximizes the mutual information of news embeddings and user embeddings in high-dimensional space to capture positive relations between news and users, optimizing the final hyperedge embeddings. Finally, the fake news classifier distinguishes the news based on integrated embeddings.

\subsection{News Semantic Channel: Updating Hyperedge Features}
When detecting fake news, news contents play a crucial role in discriminating lexical features or writing styles of fake news from real ones. Thus, we devise a news semantic channel and take the source news contents as the hyperedge attributes in the input hypergraph for our proposed model, where each hyperedge represents a piece of news. To update the textual features of news to contain semantic information, we utilize a pre-trained language model $\mathcal{M}$ to derive hyperedge features. Here we employ RoBERTa \cite{liu2020roberta} which is a robustly optimized BERT pre-trained model. The process is as follows: 
\begin{equation}
\bm{z}_{i}^{e} = \mathcal{M}(n_{i}, \forall n_{i}\in\mathcal{N}),
\label{eq:llm}
\end{equation}
where $\bm{z}^{e}_{i}$ stands for the updated hyperedge feature of a piece of news $n_{i}$. Through the language model's fine-tuning, the textual contents of news can be processed as vector embeddings that embed semantic information of news. In Hy-DeFake, since each piece of news is abstracted as a hyperedge, we take the textual embeddings as the hyperedge features $\bm{Z}^{e}$, to provide semantic information in hypergraph for accurate fake news detection results.

\subsection{User Credibility Channel: Learning Node Features}
\label{sec:user-channel}
Users are indispensable creators and disseminators of news in online social media, who are capable of making a piece of news influential by spreading it. In general, credible users are inclined to forward or propagate trustworthy news. They usually keep a wait-and-see or skeptical attitude toward uncertain or false news and do not forward them omnivorously. Oppositely, uncredible users, including potentially malicious individuals, are prone to disseminating false or misleading news. Based on this phenomenon, we suppose there exists a correlation between news and users which is worth exploring for fake news detection, thereby this work designs a user credibility channel to investigate the impact of users in identifying fake news. Since real-world social networks are complex with beyond pairwise connections, this channel learns not only user credibility features, but also the high-order correlation between news and users. 

For learning the complicated correlation between news and users with user credibility information, we propose a hypergraph Autoencoder, consisting of an encoder and a decoder in an unsupervised setting. Due to the limitation of traditional graph encoders in encoding complex correlations more than pairwise connections, Hy-DeFake adopts a hypergraph convolution network, \textit{i.e.}, HGNN \cite{feng2019hypergraph}, as the encoder. It designs a hyperedge convolution operation for learning the high-order latent correlation between nodes and hyperedges. Specifically, it performs node-edge-node transform, which can capture user-news-user relations and refine the node features of user credibility using the hypergraph structure. The hypergraph convolution is defined by
\begin{equation}
\mathbf{Z}^{(l)}=\sigma(\mathbf{D}_{v}^{-1 / 2} \mathbf{HWD}_{e}^{-1} \mathbf{H}^{\top} \mathbf{D}_{v}^{-1 / 2} \mathbf{Z}^{(l-1)}\Theta^{(l-1)}), 
\label{eq:encoder}
\end{equation}
where $\mathbf{W}$ signifies the learnable parameter during the training process, $\mathbf{Z}^{(0)} = \mathbf{X}$, $\Theta$ is the filter, and $\sigma$ indicates the nonlinear activation function.  

Through the aforementioned encoding process, the encoder maps the input data to the embeddings of users which contain user credibility information and high-order correlation between users and news. For unsupervised training of this channel, the decoder maps the embeddings back to reconstruct the input user attributes. The formulation is as follows:
\begin{equation}
\begin{split}
& \bm{\hat{x}}_{i}^{(1)} = \sigma(\Tilde{W}^{(1)}\bm{z}_{i} + b^{(1)}), \\
& \cdots \\
& \bm{\hat{x}}_{i}^{(k)} = \sigma(\Tilde{W}^{(k)}\bm{\hat{x}}_{i}^{(k-1)} + b^{(k)}).
\label{eq:decoder}
\end{split}
\end{equation}
In Equation (\ref{eq:decoder}), $\bm{z}_{i}$ represents the $i$-th node's latent representation learned by Equation (\ref{eq:encoder}), $\bm{\hat{x}}_{i}^{(k)}$ is the desired reconstructed attribute of node $i$, $\{\Tilde{W}^{(1)},\cdots,\Tilde{W}^{(k)},b^{(1)},\cdots,b^{(k)}\}$ are the parameters of the decoder with $k$ layers. The training of hypergraph Autoencoder is to minimize the reconstruction error between the input user attributes and the reconstructed user features. The loss function is as follows: 
\begin{equation}
\begin{split}
\mathcal{L}_{rec} & = \sum_{i=1}^{m}\parallel\bm{x}_{i} - \bm{\hat{x}}_{i}\parallel^{2} = \parallel\bm{X} - \bm{\hat{X}}\parallel^{2} \\
& = \parallel\bm{X} - \mathbf{D}_{v}^{-1 / 2} \mathbf{HWD}_{e}^{-1} \mathbf{H}^{\top} \mathbf{D}_{v}^{-1 / 2} \mathbf{X} \parallel^{2}. 
\label{eq:rec_loss}
\end{split}
\end{equation}

\subsection{Mutual Information-based Feature Fusion}
\label{sec:aggregation}
Through the learning process in the two channels described above, Hy-DeFake obtains the news semantic features embedded on hyperedges, the user credibility features, and high-order relations embedded on nodes in the hypergraph. Mutual information (MI) based feature fusion is proposed to integrate these semantic features, credibility features, and high-order relations to provide rich embeddings for the precise debunking of fake news. Here we first fuse the features of users who are involved in each news to gain representative user credibility embedding for each news. Then, we fuse the news semantic embeddings with the representative user credibility embeddings to obtain the integrated embeddings for fake news detection. 

Each hyperedge has multiple nodes, indicating that the news is propagated by multiple users. We define $\mathcal{U}^{s}_{i} \subseteq \mathcal{U}$ to represent a subset of users involved in news $i$. To obtain the representative embedding of users $\mathcal{U}^{s}_{i}$, we aggregate credibility embeddings of users involved in news $i$ by the element-wise mean pooling, which can be expressed as follows:   
\begin{equation}
\bm{u}_{i} = \text{MEAN}\{\bm{z}_{j}, \forall j\in\mathcal{U}^{s}_{i} \subseteq \mathcal{U}\},
\label{eq:meanpooling}
\end{equation}
where $\bm{u}_{i}$ is the representative credibility embedding of users involved in the dissemination of news $i$, and $\bm{z}_{j}$ is the credibility embedding of $j$-th user in $\mathcal{U}^{s}_{i}$. Thus, we obtain the representative user credibility matrix $\bm{U}$.  

Since we uncover a phenomenon in Section \ref{sec:user-channel}, that is users with high trustworthiness tend to share trustworthy news, whereas users with low trustworthiness are more likely to disseminate untrustworthy news. This indicates that there may be a positive correlation between news and users. Therefore, we try to maximize the mutual information between news semantic embeddings and user credibility embeddings in the training process, so as to make associated news and users closer and unassociated news and users further away in the high-dimensional space. Here we introduce a mutual information loss to capture the latent dependencies and relations between news and its relevant users who are involved in its spreading. The loss function is maximized by the following equation:
\begin{equation}
\label{eq:consis-loss}
\mathcal{L}_{mi} = \frac{1}{t}\sum_{i=1}^{t} \bigg[ \mathcal{I} (\bm{z}^{e}_{i} ; \bm{u}_{i}) + \mathcal{I}(\bm{u}_{i}) ; \bm{z}^{e}_{i}) \bigg], \\
\end{equation}
where $t$ indicates the number of news, $\bm{z}_{i}^{e}$ and $\bm{u}_{i}$ are the embedding of news $i$ and the representative embedding of users participated in spread of news $i$, respectively. The mutual information can be calculated by: 
\begin{equation}
\mathcal{I} (\bm{z}^{e}_{i} ; \bm{u}_{i}) = \mathbb{E}_{(\bm{z}_{i}^{e}, \bm{u}_{i})\sim p(\bm{z}_{i}^{e}, \bm{u}_{i})}\bigg[\log\frac{p(\bm{z}_{i}^{e}, \bm{u}_{i})}{p(\bm{z}_{i}^{e})p(\bm{u}_{i})}\bigg].
\end{equation}

Due to the assumption of positive correlation, our method maximizes the loss function in Equation (\ref{eq:consis-loss}). This maximization allows Hy-DeFake to enhance mutual information in two channels, facilitating the sharing of positive relations between credible users and real news, as well as uncredible users and fake news, respectively. Consequently, Hy-DeFake effectively learns distinctive embeddings for the task of identifying fake news.

\subsection{Fake News Detection}
After obtaining the final embeddings that fuse news semantic information and user credibility information, such integrated embeddings can be regarded as hyperedge embeddings, because each hyperedge represents an individual news piece. The task of discovering fake news becomes a hyperedge classification. In Hy-DeFake, the integrated hyperedge embeddings act as input data for the hyperedge classifier. We pass the hyperedge embeddings through a multilayer perceptron (MLP) followed by a softmax layer for final news prediction, which is formulated as follows:
\begin{gather}
\bm{z}^{o}_{i} = f(W'^{(k)} (\bm{z}_{i}^{e}\oplus\bm{u}_{i}) + b'^{(k)}), k=1,\cdots,K \\
\hat{y} = \text{SOFTMAX}(\bm{z}^{o}_{i}), 
\end{gather}
where $\bm{z}^{o}_{i}$ is the output embedding of news $i$ after MLP, $\oplus$ denotes the concatenation operation, $W'^{(k)}$ and $b'^{(k)}$ denote the parameters on layer $k$. $\hat{y}$ represents the predicted label. For each piece of news, \textit{i.e.}, each hyperedge, this component's objective is to minimize the cross-entropy loss: 
\begin{equation}
\mathcal{L}_{d} =  -y\log\hat{y}-(1-y)\log(1-\hat{y}),
\label{eq:loss_f}
\end{equation}
where $y\in\{0, 1\}$ represents the ground truth label of news. $\hat{y}$ denotes the predicted value, which indicates the probability that the news is fake. In our model Hy-DeFake, we combine the loss functions of user credibility encoding, MI-based fusion of news and users, and hyperedge classification in a unified framework. The total loss function to be optimized becomes:
\begin{equation}
\mathcal{L} = \mathcal{L}_{rec} + \mathcal{L}_{d} + \alpha\mathcal{L}_{mi},  
\label{eq:totalloss}
\end{equation}
where $\alpha$ is a coefficient to control the balance of objectives. The pseudo-code for the training of Hy-DeFake is illustrated in Algorithm \ref{alg:algorithm}.

\begin{algorithm}[!bt]
\caption{The training process of Hy-DeFake}
\label{alg:algorithm}
\textbf{Input}: $\mathcal{N}$: news content; $\mathcal{G} = (\mathcal{U}, \mathcal{E}, \bm{X})$: hypergraph of news and users; $\mathcal{Y}_{tr}$: training labels of news on hyperedges. \\
\textbf{Parameters}: $\alpha$: coefficient balancing the loss function. \\
\textbf{Output}: $\mathbf{Z}^{o}$: the learned news representations on hyperedges. 
\begin{algorithmic}[1] 
\WHILE{epoch}
\STATE Update $\bm{Z}^{e}$ of hyperedges by Eq. (\ref{eq:llm}).
\STATE Update $\bm{Z}$ of nodes by Eq. (\ref{eq:encoder}). 
\STATE Reconstruction of user attributes $\hat{\bm{X}}$ by Eq. (\ref{eq:decoder}). 
\STATE Obtain the
representative user embeddings $\bm{U}$ by Eq. (\ref{eq:meanpooling}). 
\STATE Fusion of $\bm{Z}^{e}$ and $\bm{U}$ to obtain $\bm{Z}^{o}$.
\STATE Compute the reconstruction loss $\mathcal{L}_{rec}$ by Eq. (\ref{eq:rec_loss}). 
\STATE Compute the MI loss $\mathcal{L}_{mi}$ by Eq. (\ref{eq:consis-loss}). 
\STATE Compute the cross-entropy loss $\mathcal{L}_{d}$ by Eq. (\ref{eq:loss_f}). 
\STATE Backward propagation via total loss $\mathcal{L}$ by Eq. (\ref{eq:totalloss}). \\
\ENDWHILE
\STATE \textbf{return} $\bm{Z}^{o}$ for classification. 
\end{algorithmic}
\end{algorithm}

\section{Experiments}
\label{sec:experiments}
This section validates the performance of Hy-DeFake by presenting experimental conduction and results. Specifically, after introducing the experimental setup, we assess the overall performance of Hy-DeFake by comparing it against eight popular baseline algorithms, and analyze the effectiveness of each module in Hy-DeFake, respectively. Further, we verify the explainability, robustness, and efficiency of Hy-DeFake.   

\subsection{Experimental Setup}
\subsubsection{Datasets}
\begin{table}[b]
    \centering
    \renewcommand\arraystretch{1.15}
    \caption{The Dataset Statistics.}
    \begin{tabular}{c|ccc}
    \toprule[1 pt]
    \multirow{2}{*}{\textbf{Datasets}} & \textbf{Nodes} & \textbf{Hyperedges} & \multirow{2}{*}{\textbf{Labels}} \\ 
    & \textbf{(Users)} & \textbf{(News)} & \\ \midrule
    \textbf{PolitiFact} & 27,682 & 635 & R: 344 / F: 291 \\
    \textbf{ReCOVery} & 36,741 & 2,026 & R: 1,364 / F: 662  \\
    \textbf{MM-COVID} & 18,672 & 3,992 & R: 1,894 / F: 2,098 \\
    \textbf{Gossipcop} & 72,083 & 10,084 & R: 5,444 / F: 4,640 \\
    \bottomrule[1pt]
    \end{tabular}
    \label{tab_data}
\end{table}

To investigate the efficacy of Hy-DeFake on fake news detection, we choose four real-world datasets, named PolitiFact \cite{shu2020fakenewsnet}, ReCOVery \cite{zhou2020recovery}, MM-COVID \cite{li2020mmcovid}, and Gossipcop \cite{shu2020fakenewsnet}. These datasets cover various domains, with PolitiFact focusing on political news, ReCOVery and MM-COVID containing data related to COVID-19, and Gossipcop collecting news related to entertainment. All the datasets include fact-checked source news, and the relevant social context, \textit{i.e.}, related tweets along with users who participate in the news spread. To identify fake news by incorporating the natural pattern of news outbreak and spreading, we used Twitter API\footnote{\url{https://developer.twitter.com/en/docs/twitter-api}} to crawl user information. Table \ref{tab_data} summarizes the dataset statistics, where each ``Hyperedge'' represents individual news, and ``Nodes'' represent the users participating in the news spreading. The labels assigned to the hyperedges are ``real'' and ``fake''. 

\begin{table*}[!tb]
\centering
\renewcommand\arraystretch{1.35}
\caption{Overall performance of different methods on the test datasets: Mean accuracy $\pm$ standard deviation. (D) indicates datasets, and (M) represents metrics. The best results are highlighted in gray.}
\begin{tabular}{cc|ccccccccc} 
\toprule[1pt] 
\textbf{(D)} & \textbf{(M)} & \textbf{TextCNN} & \textbf{HAN}  & \textbf{BERT} & \textbf{TextGCN} & \textbf{HyperGAT} & \textbf{DualEmo} & \textbf{HGFND} & \textbf{Llama 2} & \textbf{Hy-DeFake} \\ \midrule \midrule
\multirow{4}{*}{\rotatebox{90}{\textbf{PolitiFact}}} & \textbf{ACC} & $0.506$ \scriptsize{$\pm0.04$} & $0.521$\scriptsize{$\pm0.02$} & $0.786$\scriptsize{$\pm0.03$} & $0.74$\scriptsize{$\pm0.03$} & $0.869$\scriptsize{$\pm0.02$} & $0.83$\scriptsize{$\pm0.03$} & $0.884$\scriptsize{$\pm0.03$} & $0.913$\scriptsize{$\pm0.02$} & \cellcolor{lightgray}$0.929$\scriptsize{$\pm0.01$} \\ 
& \textbf{Pre} & $0.302$\scriptsize{$\pm0.067$} & $0.416$\scriptsize{$\pm0.05$} & $0.869$\scriptsize{$\pm0.03$} & $0.741$\scriptsize{$\pm0.03$} & $0.871$\scriptsize{$\pm0.02$} & $0.831$\scriptsize{$\pm0.03$} & $0.924$\scriptsize{$\pm0.01$} & $0.919$\scriptsize{$\pm0.02$} & \cellcolor{lightgray}$0.939$\scriptsize{$\pm0.01$} \\ 
& \textbf{Rec} & $0.529$\scriptsize{$\pm0.12$} & $0.529$\scriptsize{$\pm0.06$} & $0.808$\scriptsize{$\pm0.04$} & $0.741$\scriptsize{$\pm0.03$} & $0.869$\scriptsize{$\pm0.02$} &  $0.825$\scriptsize{$\pm0.03$} & $0.891$\scriptsize{$\pm0.05$} & $0.907$\scriptsize{$\pm0.02$} & \cellcolor{lightgray}$0.925$\scriptsize{$\pm0.01$} \\ 
& \textbf{F1} & $0.378$\scriptsize{$\pm0.07$} & $0.465$\scriptsize{$\pm0.05$} & $0.841$\scriptsize{$\pm0.02$} & $0.739$\scriptsize{$\pm0.03$} & $0.868$\scriptsize{$\pm0.02$} &  $0.824$\scriptsize{$\pm0.03$} & $0.877$\scriptsize{$\pm0.03$} & $0.91$\scriptsize{$\pm0.02$} & \cellcolor{lightgray}$0.928$\scriptsize{$\pm0.01$} \\ \hline
\multirow{4}{*}{\rotatebox{90}{\textbf{ReCOVery}}} & \textbf{ACC} & $0.406$\scriptsize{$\pm0.04$} & $0.385$\scriptsize{$\pm0.03$} & $0.756$\scriptsize{$\pm0.02$} & $0.701$\scriptsize{$\pm0.02$} & $0.648$\scriptsize{$\pm0.01$} & $0.818$\scriptsize{$\pm0.03$} & $0.923$\scriptsize{$\pm0.01$} & $0.938$\scriptsize{$\pm0.01$} & \cellcolor{lightgray}$0.952$\scriptsize{$\pm0.01$} \\ 
& \textbf{Pre} & $0.266$\scriptsize{$\pm0.06$} & $0.224$\scriptsize{$\pm0.03$} & $0.778$\scriptsize{$\pm0.02$} & $0.688$\scriptsize{$\pm0.08$} & $0.581$\scriptsize{$\pm0.01$} & $0.781$\scriptsize{$\pm0.04$} & $0.919$\scriptsize{$\pm0.01$} & $0.93$\scriptsize{$\pm0.01$} & \cellcolor{lightgray}$0.958$\scriptsize{$\pm0.01$} \\ 
& \textbf{Rec} & $0.691$\scriptsize{$\pm0.02$} & $0.675$\scriptsize{$\pm0.07$} & $0.749$\scriptsize{$\pm0.03$} & $0.616$\scriptsize{$\pm0.06$} & $0.58$\scriptsize{$\pm0.01$} & $0.763$\scriptsize{$\pm0.04$} & $0.908$\scriptsize{$\pm0.01$} & $0.922$\scriptsize{$\pm0.01$} & \cellcolor{lightgray}$0.938$\scriptsize{$\pm0.02$} \\ 
& \textbf{F1} & $0.381$\scriptsize{$\pm0.07$} & $0.336$\scriptsize{$\pm0.04$} & $0.751$\scriptsize{$\pm0.01$} & $0.597$\scriptsize{$\pm0.09$} & $0.579$\scriptsize{$\pm0.01$} & $0.769$\scriptsize{$\pm0.04$} & $0.917$\scriptsize{$\pm0.01$} & $0.926$\scriptsize{$\pm0.01$} & \cellcolor{lightgray}$0.946$\scriptsize{$\pm0.01$} \\ \hline
\multirow{4}{*}{\rotatebox{90}{\textbf{MM-COVID}}} & \textbf{ACC} & $0.496$\scriptsize{$\pm0.02$} & $0.502$\scriptsize{$\pm0.01$} & $0.879$\scriptsize{$\pm0.02$} & $0.787$\scriptsize{$\pm0.16$} & -- & $0.883$\scriptsize{$\pm0.02$} & $0.905$\scriptsize{$\pm0.01$} & $0.947$\scriptsize{$\pm0.01$} & \cellcolor{lightgray}$0.956\scriptsize{\pm0.01}$ \\ 
& \textbf{Pre} & $0.535$\scriptsize{$\pm0.02$} & $0.419$\scriptsize{$\pm0.05$} & $0.873$\scriptsize{$\pm0.02$} & $0.716$\scriptsize{$\pm0.26$} & -- & $0.881$\scriptsize{$\pm0.02$} & $0.915$\scriptsize{$\pm0.01$} & $0.945$\scriptsize{$\pm0.01$} & \cellcolor{lightgray}$0.956\scriptsize{\pm0.01}$ \\ 
& \textbf{Rec} & $0.48$\scriptsize{$\pm0.02$} & $0.48$\scriptsize{$\pm0.02$} & $0.881$\scriptsize{$\pm0.02$} & $0.744$\scriptsize{$\pm0.2$} & -- & $0.884$\scriptsize{$\pm0.02$} & $0.909$\scriptsize{$\pm0.01$} & $0.935$\scriptsize{$\pm0.01$} & \cellcolor{lightgray}$0.956\scriptsize{\pm0.01}$ \\ 
& \textbf{F1} & $0.506$\scriptsize{$\pm0.01$} & $0.446$\scriptsize{$\pm0.04$} & $0.877$\scriptsize{$\pm0.01$} & $0.696$\scriptsize{$\pm0.26$} & -- & $0.881$\scriptsize{$\pm0.02$} & $0.904$\scriptsize{$\pm0.01$} & $0.94$\scriptsize{$\pm0.01$} & \cellcolor{lightgray}$0.956\scriptsize{\pm0.01}$\\ \hline
\multirow{4}{*}{\rotatebox{90}{\textbf{Gossipcop}}} & \textbf{ACC} & $0.487$\scriptsize{$\pm0.01$} & $0.502$\scriptsize{$\pm0.01$} & $0.776$\scriptsize{$\pm0.01$} & $0.675$\scriptsize{$\pm0.11$} & $0.781$\scriptsize{$\pm0.01$} & $0.816$\scriptsize{$\pm0.01$} & $0.976$\scriptsize{$\pm0.01$} & $0.975$\scriptsize{$\pm0.01$} & \cellcolor{lightgray}$0.979\scriptsize{\pm0.01}$ \\ 
& \textbf{Pre} & $0.332$\scriptsize{$\pm0.08$} & $0.494$\scriptsize{$\pm0.04$} & $0.803$\scriptsize{$\pm0.06$} & $0.59$\scriptsize{$\pm0.21$} & $0.781$\scriptsize{$\pm0.01$} & $0.817$\scriptsize{$\pm0.01$} & $0.973$\scriptsize{$\pm0.01$} & $0.974$\scriptsize{$\pm0.01$} & \cellcolor{lightgray}$0.98\scriptsize{\pm0.01}$ \\ 
& \textbf{Rec} & $0.545$\scriptsize{$\pm0.01$} & $0.544$\scriptsize{$\pm0.01$} & $0.769$\scriptsize{$\pm0.02$} & $0.656$\scriptsize{$\pm0.13$} & $0.777$\scriptsize{$\pm0.01$} & $0.814$\scriptsize{$\pm0.01$} & \cellcolor{lightgray}$0.982$\scriptsize{$\pm0.01$} & $0.97$\scriptsize{$\pm0.01$} & $0.978$\scriptsize{$\pm0.01$} \\ 
& \textbf{F1} & $0.406$\scriptsize{$\pm0.06$} & $0.517$\scriptsize{$\pm0.02$} & $0.792$\scriptsize{$\pm0.02$} & $0.598$\scriptsize{$\pm0.2$} & $0.778$\scriptsize{$\pm0.01$} & $0.814$\scriptsize{$\pm0.01$} & $0.977$\scriptsize{$\pm0.01$} & $0.972$\scriptsize{$\pm0.01$} & \cellcolor{lightgray}$0.979$\scriptsize{$\pm0.01$} \\ 
\bottomrule[1pt]
\end{tabular}
\label{tab:results}
\end{table*}

\begin{table*}[!tb]
\centering
\renewcommand\arraystretch{1.05}
\caption{Statistical significance: Paired T-tests (The p-value less than 0.05 is significant).}
\begin{tabularx}{\textwidth}{@{\extracolsep{\fill}} cccccccccc} 
\toprule[1pt] 
\multirow{2}{*}{\textbf{Dataset}} & \multirow{2}{*}{\textbf{Metric}} & \multicolumn{8}{c}{\textbf{Hy-DeFake}} \\ \cmidrule(lr){3-10}
 & & \textbf{TextCNN} & \textbf{HAN}  & \textbf{BERT} & \textbf{TextGCN} & \textbf{HyperGAT} & \textbf{DualEmo} & \textbf{HGFND} & \textbf{Llama 2} \\ \midrule \midrule
\multirow{4}{*}{\textbf{PolitiFact}} & \textbf{ACC} & 4.26e-05 & 2.96e-06 & 0.0011 & 0.0001 & 0.0009 & 0.0027 & 0.0605 & 0.2544 \\ 
& \textbf{Pre} & 7.15e-05 & 5.51e-05 & 0.2139 & 9.45e-05 & 0.0011 & 0.0023 & 0.0289 & 0.0713 \\ 
& \textbf{Rec} & 0.0028 & 0.0002 & 0.0068 & 0.0001 & 0.0005 & 0.0032 & 0.3172 & 0.3435 \\ 
& \textbf{F1} & 0.0001 & 8.82e-05 & 0.0041 & 0.0001 & 0.0009 & 0.0024 & 0.0427 & 0.2469 \\  \hline
\multirow{4}{*}{\textbf{ReCOVery}} & \textbf{ACC} & 1.49e-05 & 6.62e-06 & 0.0003 & 2.55e-05 &  5.92e-07 & 0.0025 & 0.0022 & 0.0554 \\  
& \textbf{Pre} & 2.11e-05 & 2.74e-06 & 0.005 & 0.0024 & 3.61e-07 & 0.0013 & 0.0029 & 0.0258 \\ 
& \textbf{Rec} & 4.82e-05 & 0.0029 & 0.0113 & 0.0007 & 4.84e-06 & 0.0029 & 0.0236 & 0.1885 \\ 
& \textbf{F1} & 7.42e-05 & 2.01e-05 & 0.0005 & 0.002 & 1.76e-06 & 0.0017 & 0.0096 & 0.0832 \\ \hline
\multirow{4}{*}{\textbf{MM-COVID}} & \textbf{ACC} & 1.01e-06 & 4.92e-08 & 0.0001 & 0.0007 & -- & 0.0012 & 0.0015 & 0.1646  \\ 
& \textbf{Pre} & 8.77e-07 & 4.49e-05 & 0.0004 & 0.0005 & -- & 0.002 & 0.0007 & 0.1151  \\ 
& \textbf{Rec} & 8.55e-07 & 1.08e-06 & 5.06e-06 & 0.0002 & -- & 0.0011 & 0.0017 & 0.0374  \\ 
& \textbf{F1} & 2.97e-07 & 1.39e-05 & 7.95e-05 & 0.0002 & -- & 0.0014 & 0.0015 & 0.0751  \\  \hline
\multirow{4}{*}{\textbf{Gossipcop}} & \textbf{ACC} & 5.00e-08 & 6.80e-10 & 5.37e-08 & 0.0059 & 1.09e-07 & 2.23e-07 & 0.5771 & 0.5903 \\ 
& \textbf{Pre} & 8.84e-05 & 1.07e-05 & 0.0004 & 0.0225 & 1.56e-07 & 9.58e-08 & 0.1183 & 0.4594 \\ 
& \textbf{Rec} & 6.68e-07 & 9.52e-09 & 7.67e-05 & 0.0084 & 8.14e-08 & 1.84e-06 & 0.5589 & 0.3885 \\ 
& \textbf{F1} & 4.10e-05 & 1.15e-06 & 1.06e-05 & 0.0203 & 9.14e-08 & 7.74e-07 & 0.5913 & 0.4197 \\ 
\bottomrule[1pt]
\end{tabularx}
\label{tab:results_pvalue}
\end{table*}

\subsubsection{Baselines and Metrics} 
In our evaluation, we employ four frequently employed metrics to compare the performance of our method with eight popular baseline algorithms. Here is a brief introduction to the baseline algorithms: 
\begin{itemize}
\item TextCNN \cite{kim-2014-convolutional} is a CNN (convolutional neural network) based method to classify sentences, it builds CNNs and utilizes convolution filters on \textit{word2vec} of news contents to capture textual features in different granularity. 
\item HAN \cite{yang2016hierarchical} (hierarchical attention network) is proposed to classify documental content. It adopts an attention mechanism at both the sentence and word levels to encode news content. 
\item BERT \cite{devlin2018bert} is a prominent pre-trained language model for natural language processing (NLP) tasks, that utilizes a transformer-based architecture with a bidirectional encoder and self-attention heads. Here we fine-tune the BERT model specifically for identifying fake news. 
\item TextGCN \cite{yao2019graph} is a graph-based approach that extends GCN \cite{kipf2017semisupervised} to perform text classification. It captures semantic relationships between words and sentences in the graph representation.  
\item HyperGAT \cite{ding2020more} is a hypergraph-based inductive method for textual data. It constructs a document-level hypergraph attention network for each piece of news to learn the textual representations in fake news detection. 
\item DualEmo \cite{dualemo2021zhang} is the fake news detection model that mines dual emotions from publishers and users in a social context. It identifies the emotional gap between source news and relevant comments to distinguish fake news. 
\item HGFND \cite{jeong2022nothing} is a hypergraph learning-based method to capture group-wise interactions among news. It leverages a dual-level attention mechanism to represent important news relations.  
\item Llama 2 \cite{touvron2023llama} is a collection of fine-tuned and pre-trained large language models (LLMs). Its parameters range in scale from 7 billion to 70 billion. Here we utilize the chat model to learn the representations of news and then identify real and fake news by classifier.   
\end{itemize} 

In this paper, we measure the performance of our proposed Hy-DeFake and eight baseline methods for the detection of fake news using four widely-used metrics: Accuracy (ACC), Precision (Pre), Recall (Rec), and F1 score (F1). 

\subsubsection{Experimental Setup and Implementation}
The datasets are split into train-validation-test sets with a ratio of 70\%-10\%-20\%. In Hy-DeFake, the embedding size is set to 768. We use 2 hypergraph convolution layers and 3 fully-connected layers in both the decoder and hyperedge classification. The model optimization is performed using the Adam optimizer with a learning rate of 0.001 for all datasets. The number of epochs is set to 600. The evaluation is conducted using 5-fold cross-validation.

\begin{table*}[!htb]
\centering
\renewcommand\arraystretch{1.1}
\caption{Results of ablation study on test datasets (Mean $\pm$ Stdv, best in gray).}
\begin{tabularx}{\textwidth}{@{\extracolsep{\fill}} ccccccccc} 
\toprule[1pt] 
\multirow{2}{*}{\textbf{Variants}} & \multicolumn{4}{c}{\textbf{PolitiFact}} & \multicolumn{4}{c}{\textbf{ReCOVery}} \\ \cmidrule(lr){2-5}\cmidrule(lr){6-9}
 & \textbf{ACC} & \textbf{Pre}  & \textbf{Rec} & \textbf{F1} & \textbf{ACC} & \textbf{Pre} & \textbf{Rec} & \textbf{F1} \\ \midrule
 \textbf{\textit{w} Sem} & $0.81$ \scriptsize{$\pm0.01$} & $0.815$ \scriptsize{$\pm0.01$} & $0.814$ \scriptsize{$\pm0.01$} & $0.807$ \scriptsize{$\pm0.01$} & $0.759$ \scriptsize{$\pm0.04$} & $0.744$ \scriptsize{$\pm0.04$} & $0.733$ \scriptsize{$\pm0.06$} & $0.721$ \scriptsize{$\pm0.05$} \\ 
\textbf{\textit{w} Cre} & $0.521$ \scriptsize{$\pm0.04$} & $0.394$ \scriptsize{$\pm0.19$} & $0.501$ \scriptsize{$\pm0.01$} & $0.353$ \scriptsize{$\pm0.02$} & $0.666$ \scriptsize{$\pm0.02$} & $0.649$ \scriptsize{$\pm0.18$} & $0.503$ \scriptsize{$\pm0.01$} & $0.406$ \scriptsize{$\pm0.01$} \\ 
\textbf{Sem + Cre} & $0.909$ \scriptsize{$\pm0.01$} & $0.915$ \scriptsize{$\pm0.01$} & $0.908$ \scriptsize{$\pm0.01$} & $0.908$ \scriptsize{$\pm0.01$} & $0.921$ \scriptsize{$\pm0.02$} & $0.922$ \scriptsize{$\pm0.01$} & $0.899$ \scriptsize{$\pm0.02$} & $0.908$ \scriptsize{$\pm0.01$} \\ 
\textbf{Hy-DeFake} & \cellcolor{lightgray}$0.929$ \scriptsize{$\pm0.01$} & \cellcolor{lightgray}$0.939$ \scriptsize{$\pm0.01$} & \cellcolor{lightgray}$0.925$ \scriptsize{$\pm0.01$} & \cellcolor{lightgray}$0.928$ \scriptsize{$\pm0.01$} & \cellcolor{lightgray}$0.952$ \scriptsize{$\pm0.01$} & \cellcolor{lightgray}$0.958$ \scriptsize{$\pm0.01$} & \cellcolor{lightgray}$0.938$ \scriptsize{$\pm0.02$} & \cellcolor{lightgray}$0.946$ \scriptsize{$\pm0.01$} \\  \hline \hline
\multirow{2}{*}{\textbf{Variants}} & \multicolumn{4}{c}{\textbf{MM-COVID}} & \multicolumn{4}{c}{\textbf{Gossipcop}} \\ \cmidrule(lr){2-5}\cmidrule(lr){6-9}
 & \textbf{ACC} & \textbf{Pre}  & \textbf{Rec} & \textbf{F1} & \textbf{ACC} & \textbf{Pre} & \textbf{Rec} & \textbf{F1} \\ \midrule
 \textbf{\textit{w} Sem} & $0.853$ \scriptsize{$\pm0.03$} & $0.879$ \scriptsize{$\pm0.02$} & $0.859$ \scriptsize{$\pm0.02$} & $0.851$ \scriptsize{$\pm0.03$} & $0.797$ \scriptsize{$\pm0.04$} & $0.834$ \scriptsize{$\pm0.03$} & $0.79$ \scriptsize{$\pm0.05$} & $0.786$ \scriptsize{$\pm0.05$} \\ 
\textbf{\textit{w} Cre} & $0.486$ \scriptsize{$\pm0.03$} & $0.424$ \scriptsize{$\pm0.11$} & $0.492$ \scriptsize{$\pm0.03$} & $0.397$ \scriptsize{$\pm0.06$} & $0.469$ \scriptsize{$\pm0.1$} & $0.394$ \scriptsize{$\pm0.18$} & $0.46$ \scriptsize{$\pm0.08$} & $0.319$ \scriptsize{$\pm0.05$} \\ 
\textbf{Sem + Cre} & $0.929$ \scriptsize{$\pm0.01$} & $0.935$ \scriptsize{$\pm0.01$} & $0.929$ \scriptsize{$\pm0.01$} & $0.928$ \scriptsize{$\pm0.01$} & $0.959$ \scriptsize{$\pm0.01$} & $0.961$ \scriptsize{$\pm0.01$} & $0.961$ \scriptsize{$\pm0.01$} & $0.959$ \scriptsize{$\pm0.01$} \\ 
\textbf{Hy-DeFake} & \cellcolor{lightgray}$0.956$ \scriptsize{$\pm0.01$} & \cellcolor{lightgray}$0.956$ \scriptsize{$\pm0.01$} & \cellcolor{lightgray}$0.956$ \scriptsize{$\pm0.01$} & \cellcolor{lightgray}$0.956$ \scriptsize{$\pm0.01$} & \cellcolor{lightgray}$0.979$ \scriptsize{$\pm0.01$} & \cellcolor{lightgray}$0.98$ \scriptsize{$\pm0.01$} & \cellcolor{lightgray}$0.978$ \scriptsize{$\pm0.01$} & \cellcolor{lightgray}$0.979$ \scriptsize{$\pm0.01$} \\ 
\bottomrule[1pt]
\end{tabularx}
\label{tab_ablation}
\end{table*}

\subsection{Performance}
The performance of Hy-DeFake, along with eight popular baseline methods, is evaluated on four real-world datasets. The results are presented in TABLE \ref{tab:results}, with the optimal results highlighted in gray. As shown in this table, the performance of Hy-DeFake is superior to that of the baseline methods on three datasets, and achieves sub-optimal or optimal performance on one dataset, \textit{i.e.}, Gossipcop. This illustrates that the embeddings of Hy-DeFake that incorporate news semantics and user credibility learn sufficient information compared to other methods. Further, capturing high-order correlation between news and users is effective for identifying fake news. To analyze the statistical significance of Hy-DeFake's improvement compared with baseline methods, we conducted paired T-tests and present the results of p-values in TABLE \ref{tab:results_pvalue}, where most of the p-values are less than 0.05, indicating significant differences. 

Regarding the baseline performances, 
Llama 2 exhibits sub-optimal performance, as it is a large language model with billions of parameters. The version of Llama 2 utilized in our experiments is the chatbot model, which conducts text generation inference and demonstrates a heightened understanding of semantics in news content. The performance of HGFND is also satisfactory, as it considers high-order relations in social networks as well. However, it still relies on vanilla GNNs to learn hypergraph representations and is limited in capturing sufficient high-order relations. Moreover, HGFND requires numerous types of hyperedges when extracting hypergraphs, increasing the difficulty of dataset construction and impacting real-world practicality. The overall results of DualEmo are also notable, because this method considers social context as well when detecting fake news. It verifies that considering social context \textit{e.g.}, dual emotion between publishers and users, is valuable for distinguishing between real and fake news. 

HyperGAT is the other baseline method that uses the concept of hypergraph, but it builds hypergraphs of words in a piece of news, which leads to worse performance than our method due to ignoring user information. Moreover, the construction of hypergraphs makes HyperGAT require the length of text. A hypergraph cannot be built if the text is too short. However, most COVID-19 news is in the form of short tweets, such as those in the MM-COVID dataset, so HyperGAT does not work on this dataset (mark as ``--'' in TABLE \ref{tab:results}). The reason why HyperGAT can perform on the other three datasets is that the datasets provide statements for each news. Thus, we incorporate these textual data to serve as supplementary information and make the data meet the requirements of HyperGAT. TextGCN builds graphs at the sentence level, so that the structural information of the sentence can be taken into account. Nevertheless, short news in tweets fails to learn enriched information about the structure among sentences, resulting in mediocre results. 

The aforementioned four baseline methods (\textit{i.e.}, HGFND, DualEmo, HyperGAT, and TextGCN) take into account structural or social context information, while the rest of the baseline methods only consider textual information from news content. Apart from Llama 2, another baseline algorithm that performs stable is BERT, indicating that BERT can learn effective textual features of news. However, BERT's performance in fake news detection is mediocre and obviously worse than that of Hy-DeFake and other social context-based baselines, which reflects that only learning textual information and ignoring the assistance of social context like user information leads to limited results. In TextCNN and HAN methods, because they do not learn structural information, are not pre-trained in the large-scale corpus, and do not take social contexts into account, they perform the worst. 

Comparing the overall results in TABLE \ref{tab:results} and the statistical significance in TABLE \ref{tab:results_pvalue}, it shows that the methods that consider side information perform better than methods that only use textual information. This demonstrates that social context and structural information are indeed instrumental in detecting fake news. Overall, Hy-DeFake consistently delivers accurate and reliable outcomes for distinguishing fake news in various domains.

\subsection{Ablation Study}
\label{sec:ablation}
To assess the impact of the key components in Hy-DeFake, ablation experiments are conducted to evaluate the performance of the news semantic channel, user credibility channel, and MI-based feature fusion individually. The statistical outcomes of these ablation studies are presented in TABLE \ref{tab_ablation}, where ``\textit{w} Sem'' and ``\textit{w} Cre'' indicate that the news semantic channel and user credibility channel are solely used, respectively. ``Sem + Cre'' denotes the model incorporating the news semantic channel and the user credibility channel, while excluding MI-based feature fusion. ``Hy-DeFake'' is the final model with all components. Fig. 5 illustrates that all evaluation metrics on four databases are significantly increased when the user or news signals are not solely used to train our model. 

The ablation studies indicate that the performance of ``\textit{w} Cre'' is the poorest. That is because the model solely learns user credibility and neglects the objective of classifying news during training. In this case, user attributes play a limited role and the textual information of news is absent to accomplish accurate detection of fake news. This underscores that Hy-DeFake remains text-dependent; relying solely on user information cannot accomplish effective fake news detection. 

When relying solely on textual information from news content, the outcomes of ``\textit{w} Sem'' are similar to those obtained by text-based language models. This is because they exclusively consider semantic information and disregard the significance of social context in detecting fake news. However, when considering user information simultaneously, the direct concatenation of news embeddings and user embeddings in ``Sem + Cre'' has a significantly improved effect compared to that of ``\textit{w} Sem''. This implies that incorporating social contexts is crucial for superior performance. By introducing MI-based feature fusion, our final model Hy-DeFake exhibits further improvement. This indicates that imposing mutual information between news and users in a high-dimensional space can accentuate the distinction between trustworthy and untrustworthy users as well as real and fake news, thereby rendering the learned representations more discriminative. 

To conclude, each individual component has a significant contribution to the proposed Hy-DeFake for detecting fake news in online social networks.

\begin{figure*}[!htb]
\centering
\subfloat[PolitiFact]{\includegraphics[width=0.4\textwidth]{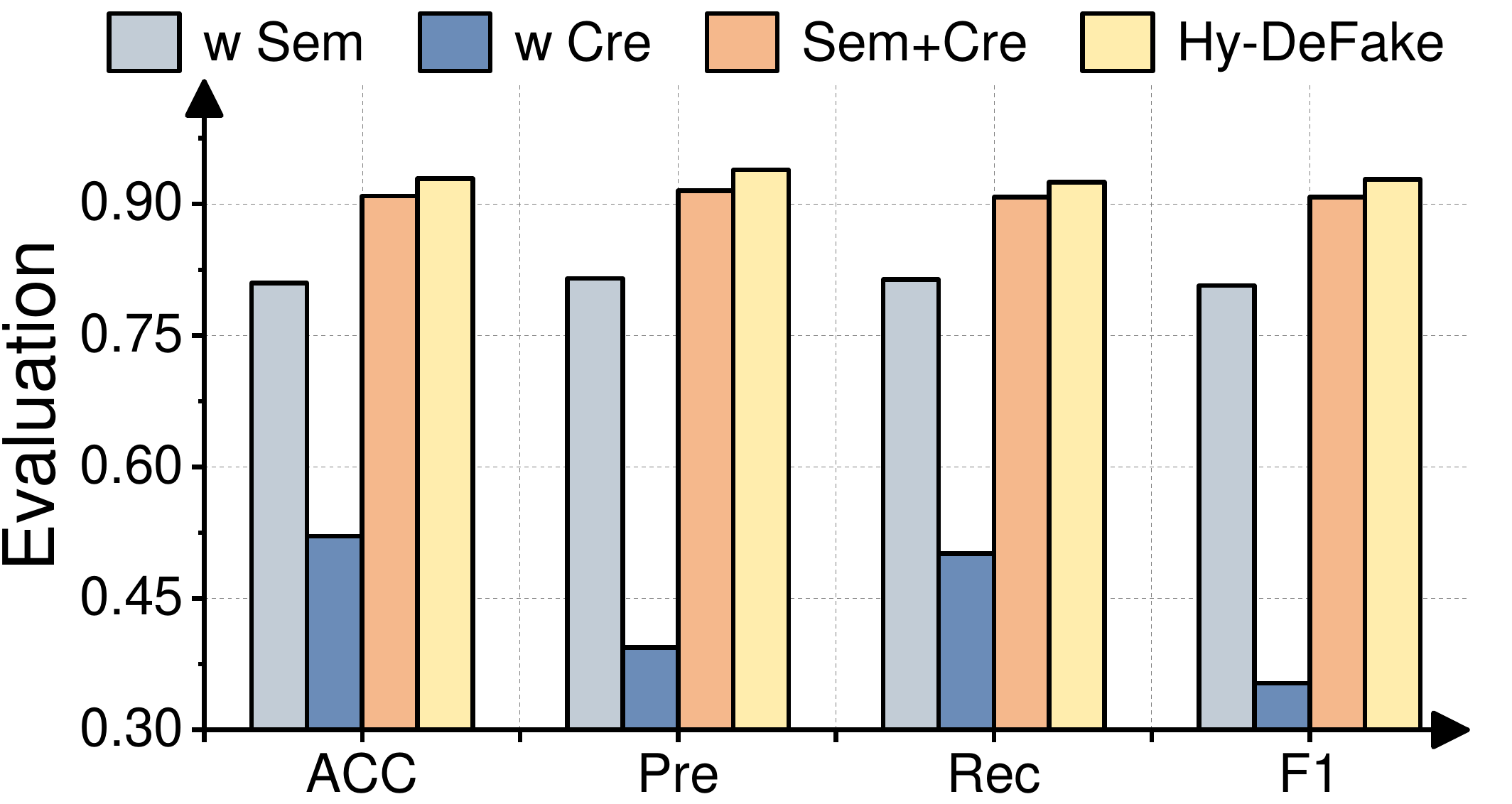}}
\subfloat[ReCOVery]{\includegraphics[width=0.4\textwidth]{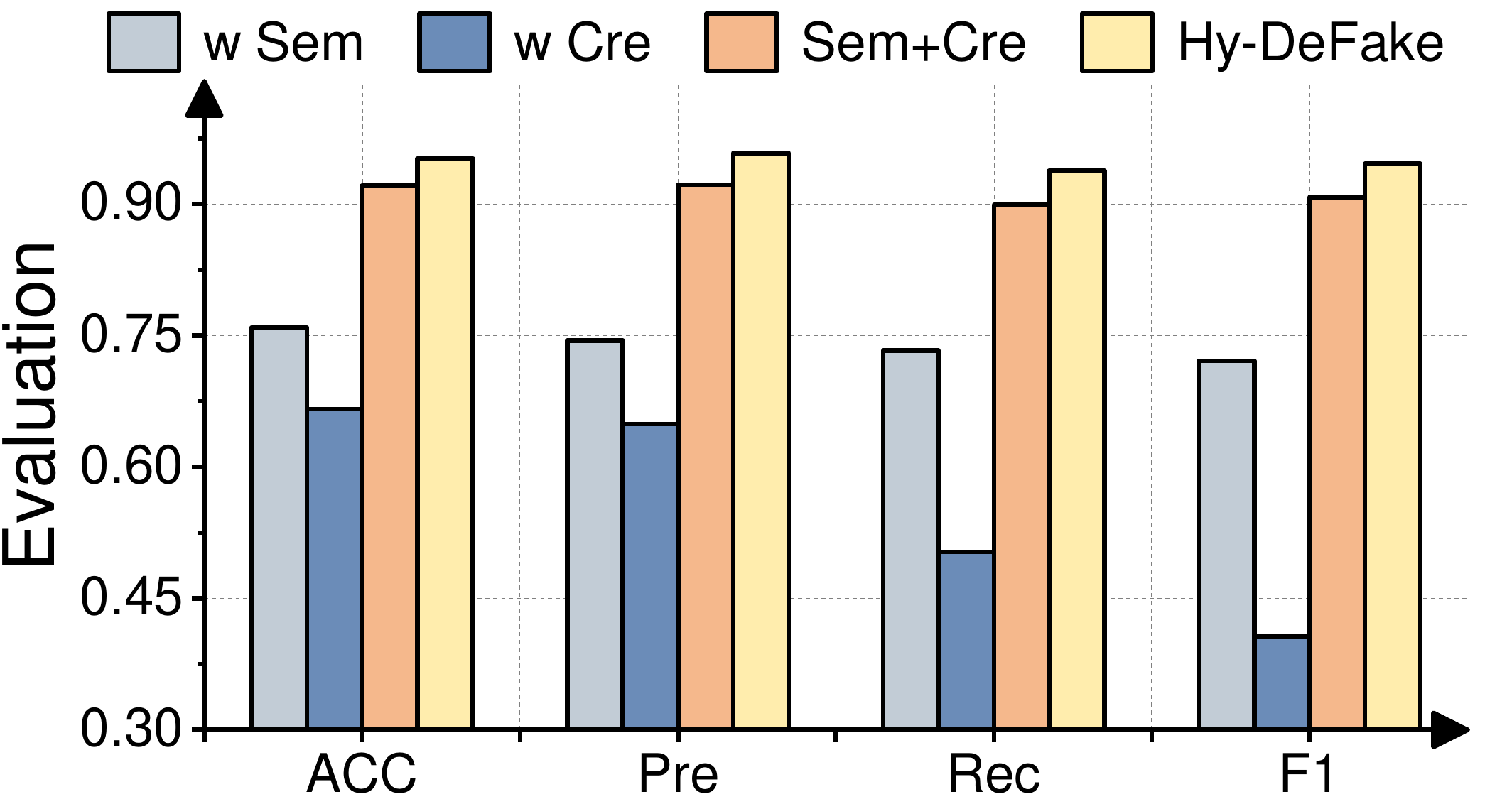}}\\
\subfloat[MM-COVID]{\includegraphics[width=0.4\textwidth]{abla_recovery.pdf}}
\subfloat[Gossipcop]{\includegraphics[width=0.4\textwidth]{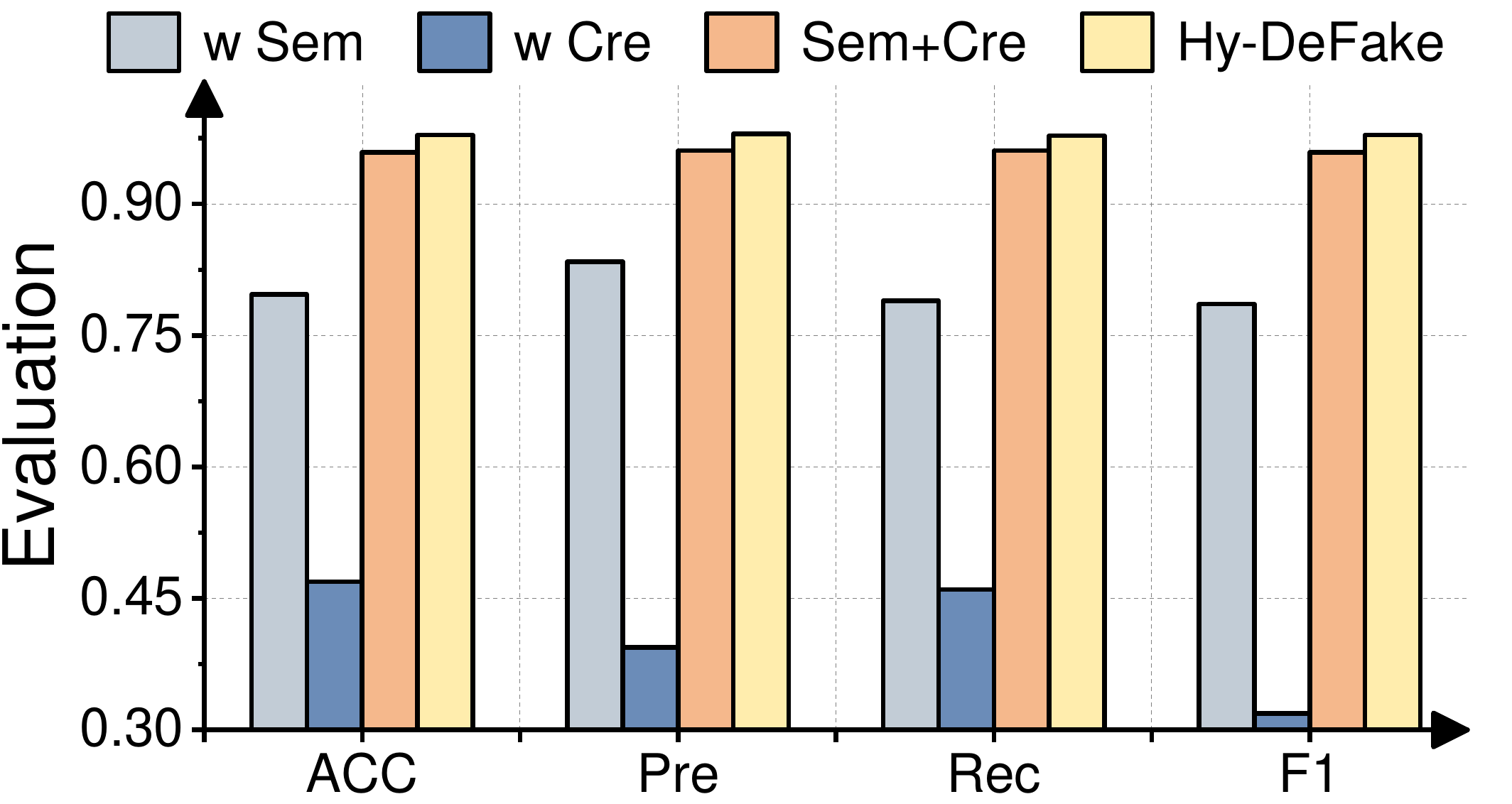}}
\caption{Experimental results of ablation study on four real-world datasets. ``\textit{w} Sem'' trains news semantic channel solely, ``\textit{w} Cre'' trains user credibility channel individually, ``Sem + Cre'' integrates news embeddings and user embeddings without MI-based fusion, and ``Hy-DeFake'' is the proposed model. }
\label{fig:ablation}
\end{figure*}

\subsection{Case Study and Visualization}
\label{sec:casestudy}
To examine the influence of user credibility on news reliability, we choose the PolitiFact dataset to perform a case study. Here we analyze the distinctions between users who are involved in spreading fake news versus real news.

\begin{table}[htb]
\renewcommand\arraystretch{1.25}
\setlength{\tabcolsep}{1.5mm}
    \centering
    \begin{tabular}{c|cc}
    \toprule
        \textbf{Statistics} & \textbf{Real News} & \textbf{Fake News}  \\
    \midrule
        Average number of users in each news & 43 & 46 \\ \hline
        Average number of following of involved users & 2,555 & 3,468 \\ \hline
        Average number of followers of involved users & 48,150 & 9,140 \\ \hline 
        Verified Users & 1,053 & 167 \\ \hline 
        Total number of users in all news & 14,822 & 13,371 \\   
    \bottomrule    
    \end{tabular}
    \caption{User Statistics in PolitiFact Dataset.}
    \label{tab:casestudy}
\end{table}

We analyze the disparities in user attributes between real and fake news by presenting key statistics of users involved in news spreading in the PolitiFact dataset, as displayed in TABLE \ref{tab:casestudy}. The average number of users in each piece of news demonstrates that fake news tends to attract more attention compared to real news. For real news, the average number of followers significantly exceeds the average number of following, indicating that the users involved in real news are influential and potentially represent public accounts. However, such discrepancy is less obvious in fake news. Furthermore, the number of verified users in real news is nearly 10 times more than that in fake news. This attribute reflecting user credibility is positively correlated with news authority. By comparing the total number of users in all news and the average number of users in each news, we observe that the total number of users involved in fake news is less than that in real news, but the average number of users in each fake news surpasses that in real news. It illustrates that users involved in fake news spreading are more active, and some may be malicious users who spread false news in large numbers. 

\begin{figure}[!htbp]
\centering
\includegraphics[scale=0.45]{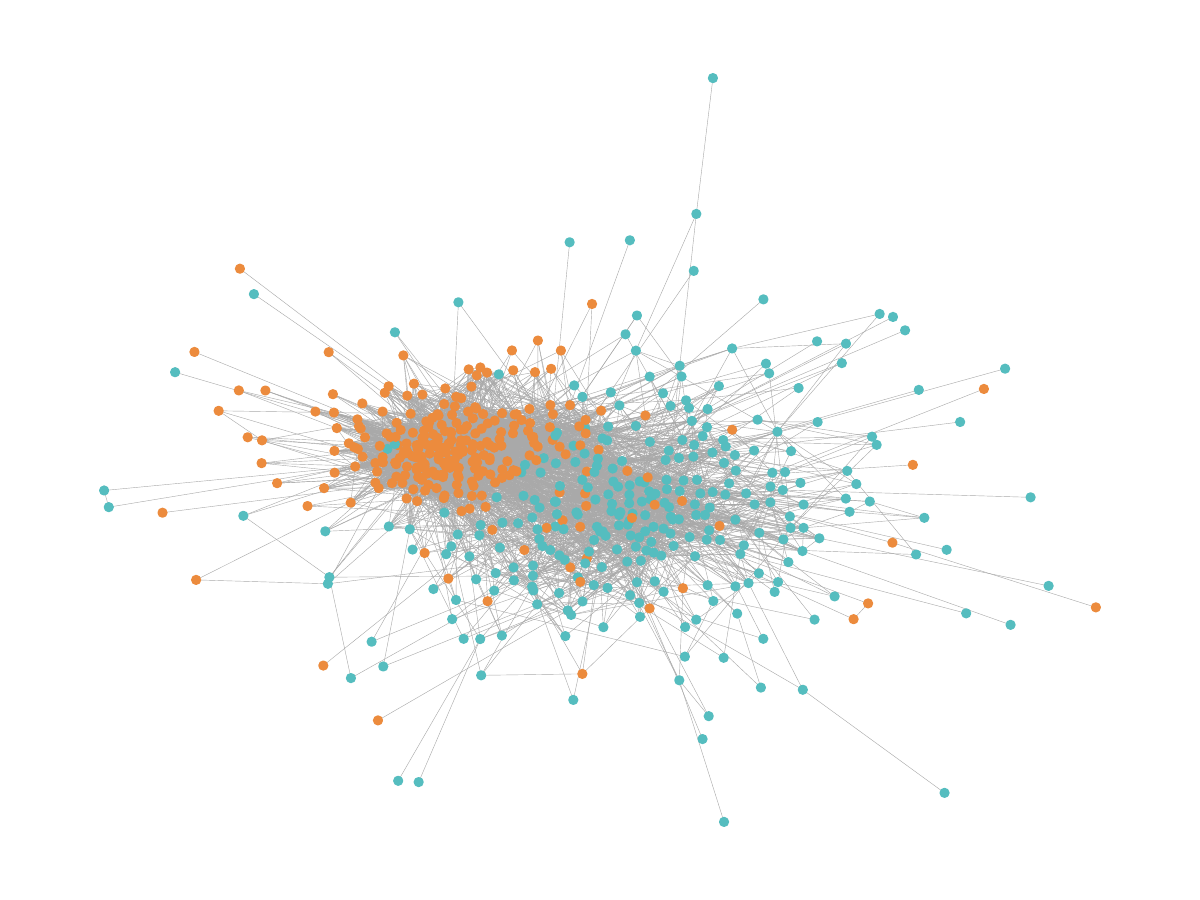} 
\caption{User relation when spreading news in PolitiFact dataset (Blue nodes: users who spread real news; Orange nodes: users who spread fake news; Edges: connect any two users who participate in the same piece of news). }
\label{fig:casestudy}
\end{figure}

In addition to the above analysis of the relationship between news and users from the perspective of statistics and attributes, we also analyze from the perspective of structure in Fig. \ref{fig:casestudy}. 
In this figure, the orange nodes represent the users involved in fake news while the blue nodes stand for the users who spread real news. The edges link two users who are involved in the same piece of news. We discover that users who spread fake news connect more closely while those involved in real news are evenly distributed. The users spreading fake news form a dense community, as some malicious users deliberately participate in the spread of multiple fake news. Therefore, Fig. \ref{fig:casestudy} demonstrated that exploring structural information in the social context is essential for fake news detection.  

\begin{figure}[!htb]
\centering
\subfloat[Original]{\includegraphics[width=0.35\textwidth]{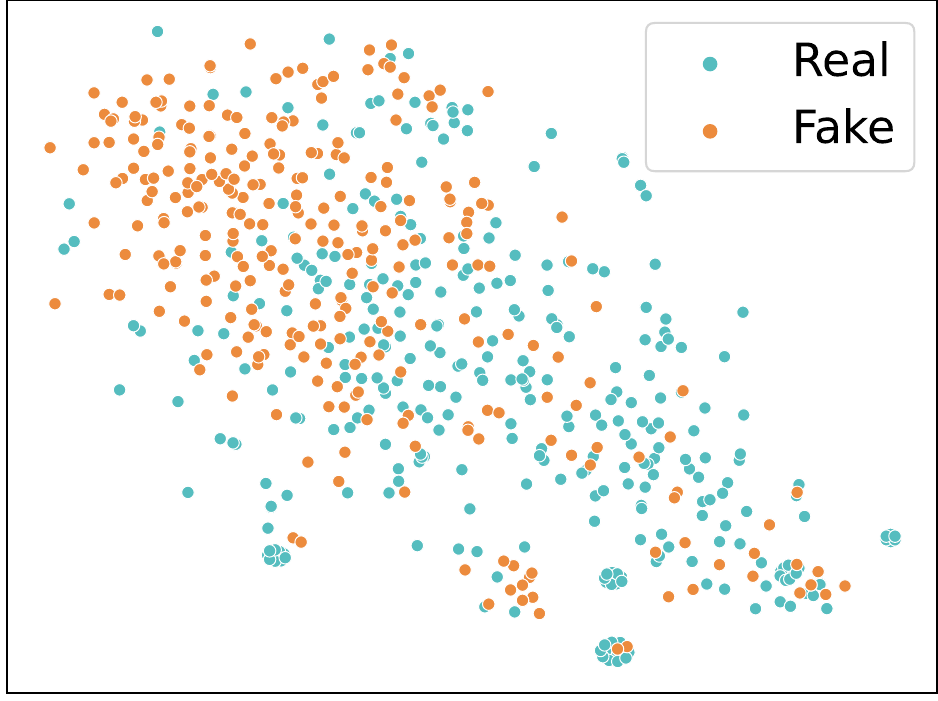}}\\
\subfloat[Representations]{\includegraphics[width=0.35\textwidth]{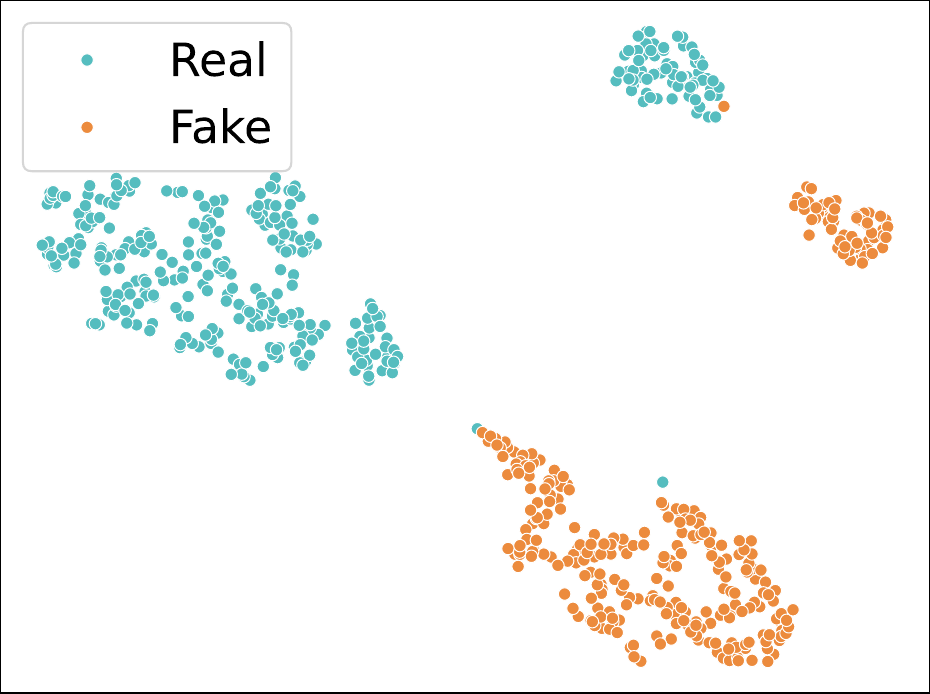}}
\caption{t-SNE visualization of PolitiFact dataset.}
\label{fig:tsne}
\end{figure}

Finally, to visually observe the representations learned by Hy-DeFake, we applied t-SNE technology to project the high-dimensional representations of the PolitiFact dataset into a 2D space, as shown in Fig. \ref{fig:tsne}. Comparing the distribution with the original data, Hy-DeFake distinctly projects representations of real and fake news into different regions, demonstrating the effectiveness of our method.

\subsection{Fake News Mitigation and Early Detection}

\begin{figure}[!htb]
\centering
\subfloat[Performance]{\includegraphics[width=0.42\textwidth]{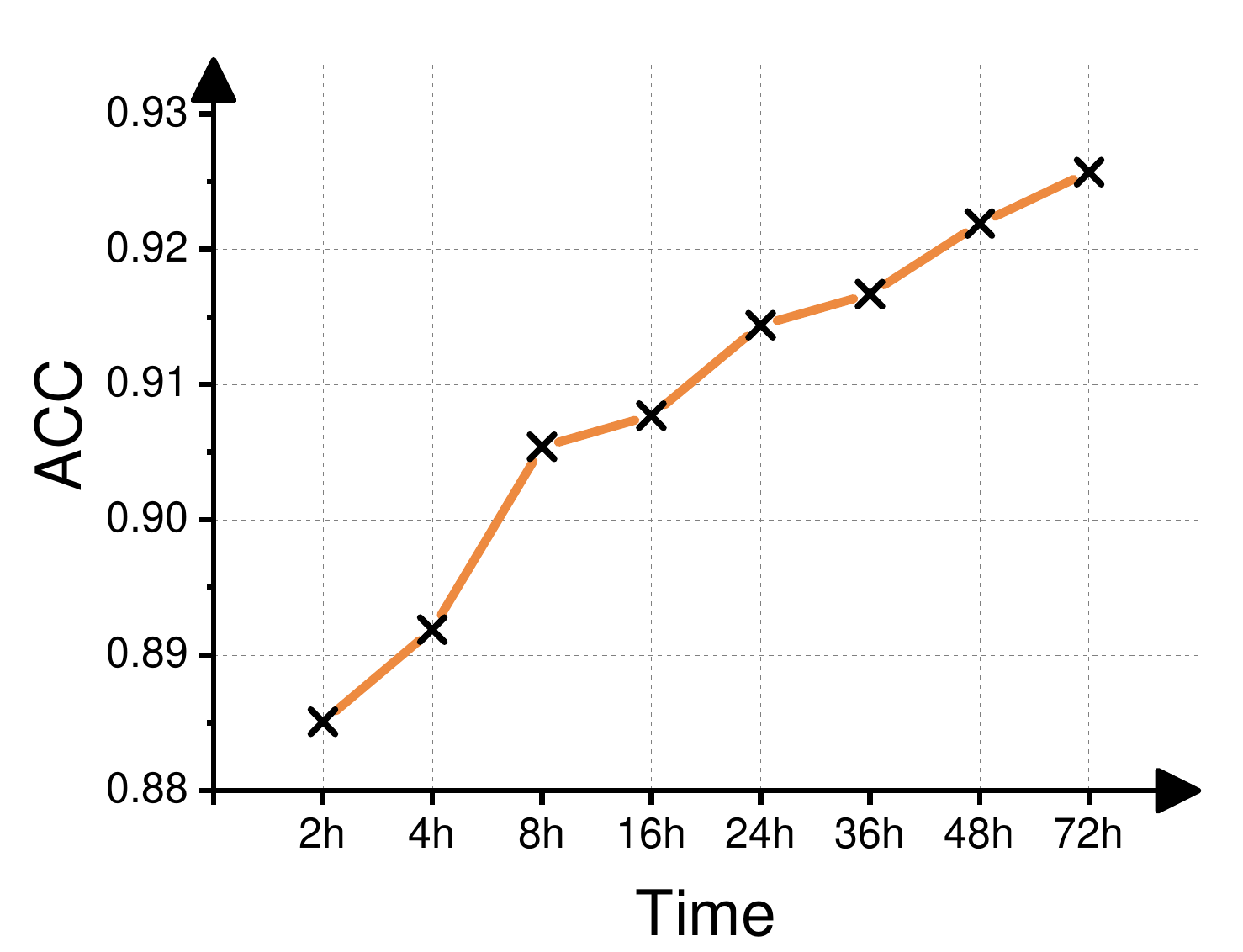}}\\
\subfloat[Participated Users]{\includegraphics[width=0.4\textwidth]{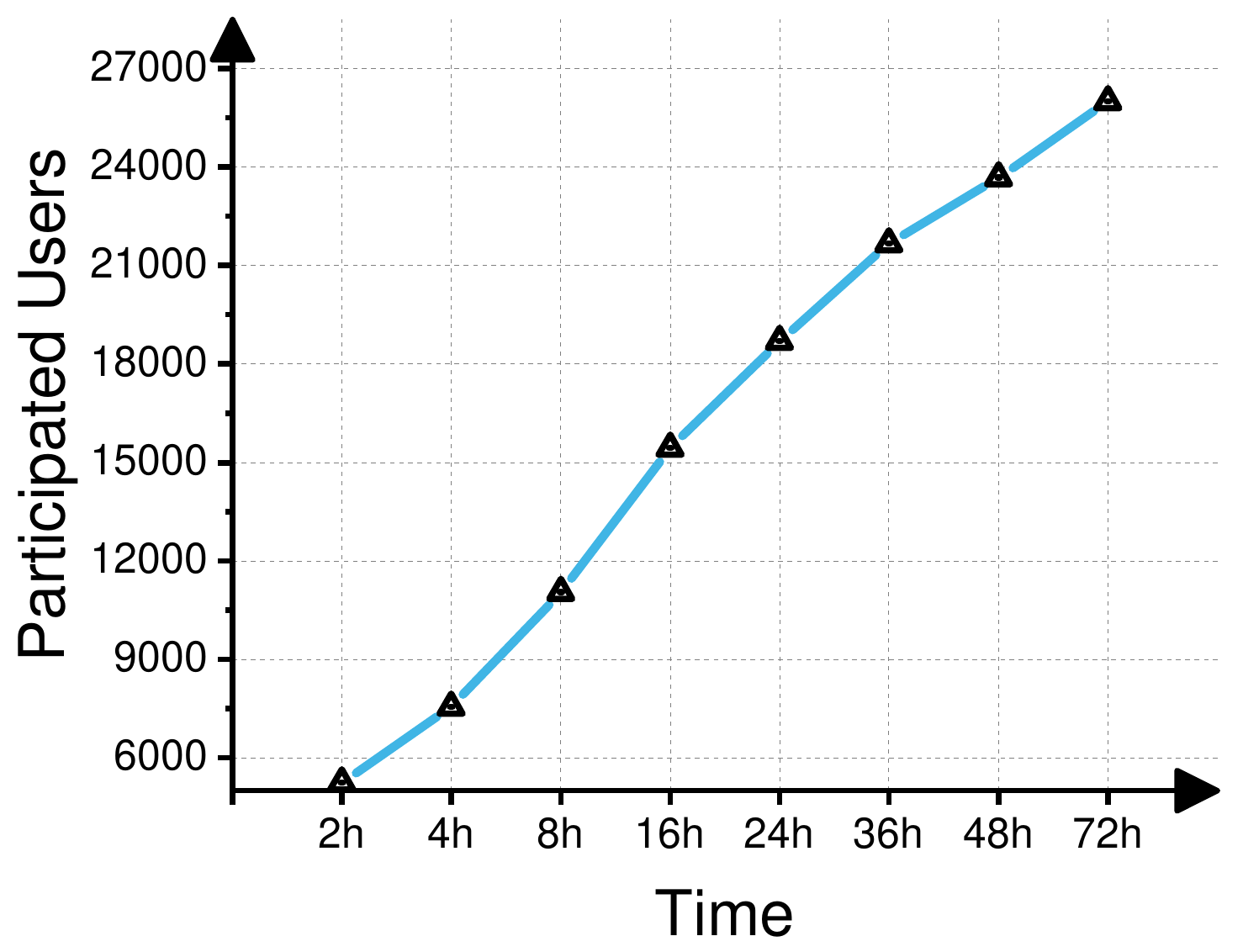}}
\caption{Results of fake news early detection during news spreading on PolitiFact dataset. \textbf{(a) Performance} illustrates the variation of Hy-DeFake's accuracy over time, while \textbf{(b) Participated Users} depicts the evolution of users engaged in the dissemination of news over time. } 
\label{fig:early}
\end{figure}

Detecting and mitigating the spread of fake news before it causes harm is crucial. Therefore, we conducted early detection experiments on the PolitiFact dataset to assess the applicability of Hy-DeFake in news spreading. The results are presented in Fig. \ref{fig:early}. In the experimental settings, we evaluate the performance of Hy-DeFake at intervals of 2 hours, 4 hours, 8 hours, 16 hours, 24 hours, 36 hours, 48 hours, and 72 hours after the news publication. Notably, Hy-DeFake exhibits satisfactory performance within the first 2 hours, and achieves above 90 \% accuracy after 4 hours. The experimental results validate the effectiveness of our method in detecting fake news at the early stages of news propagation, owing to its ability to capture higher-order relations in the social context.

\subsection{Parameter Sensitivity}
\begin{figure*}[!htb]
\centering
\subfloat[PolitiFact]{\includegraphics[width=0.25\textwidth]{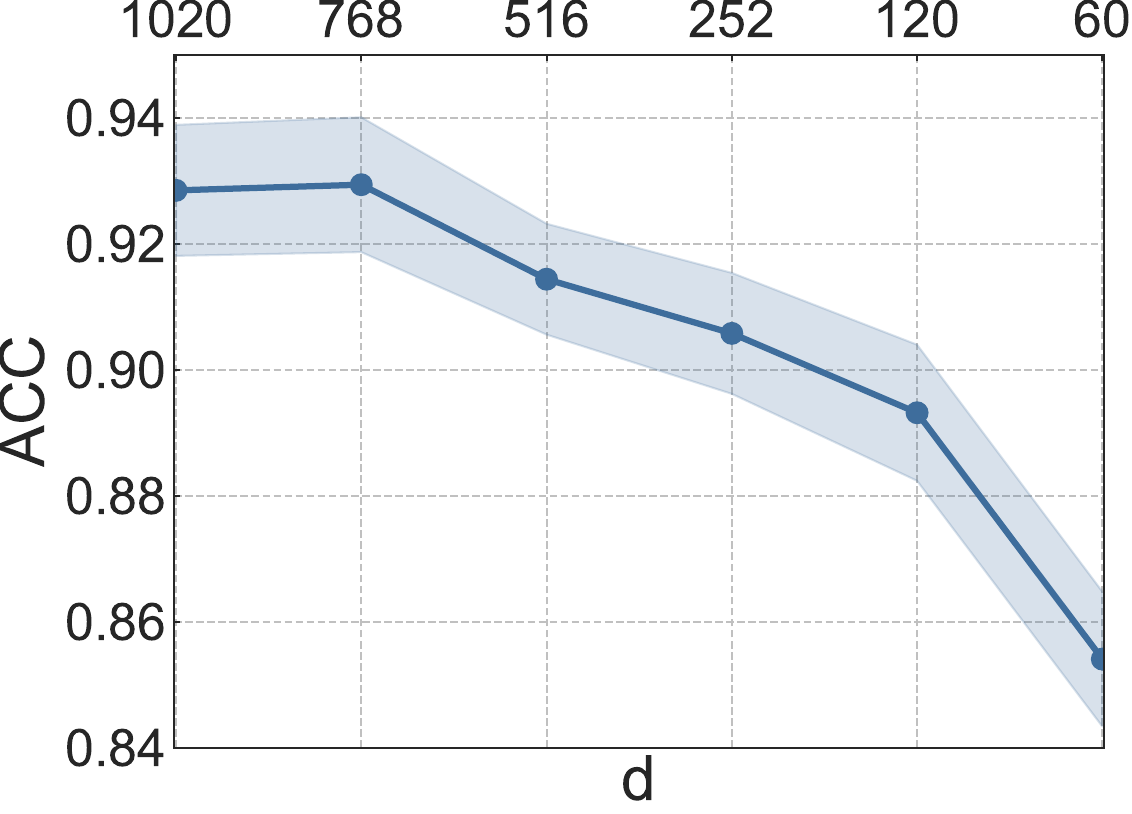}}
\subfloat[ReCOVery]{\includegraphics[width=0.25\textwidth]{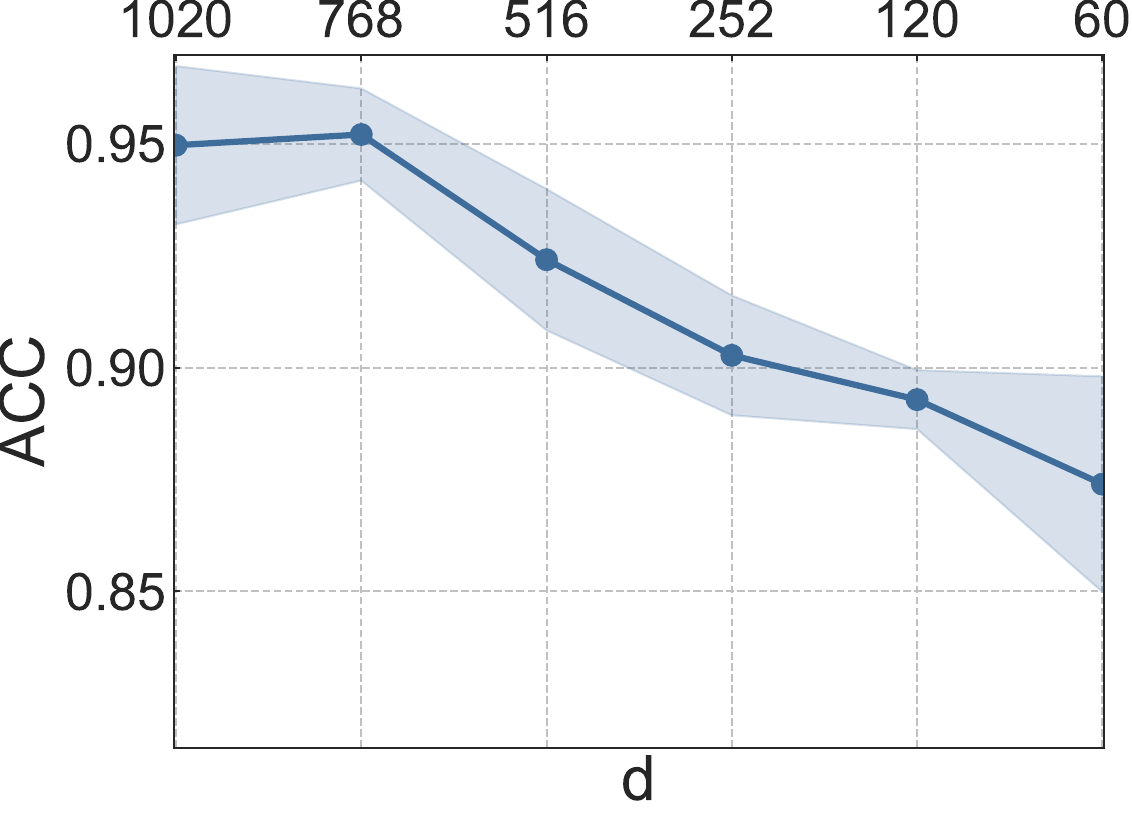}}
\subfloat[MM-COVID]{\includegraphics[width=0.25\textwidth]{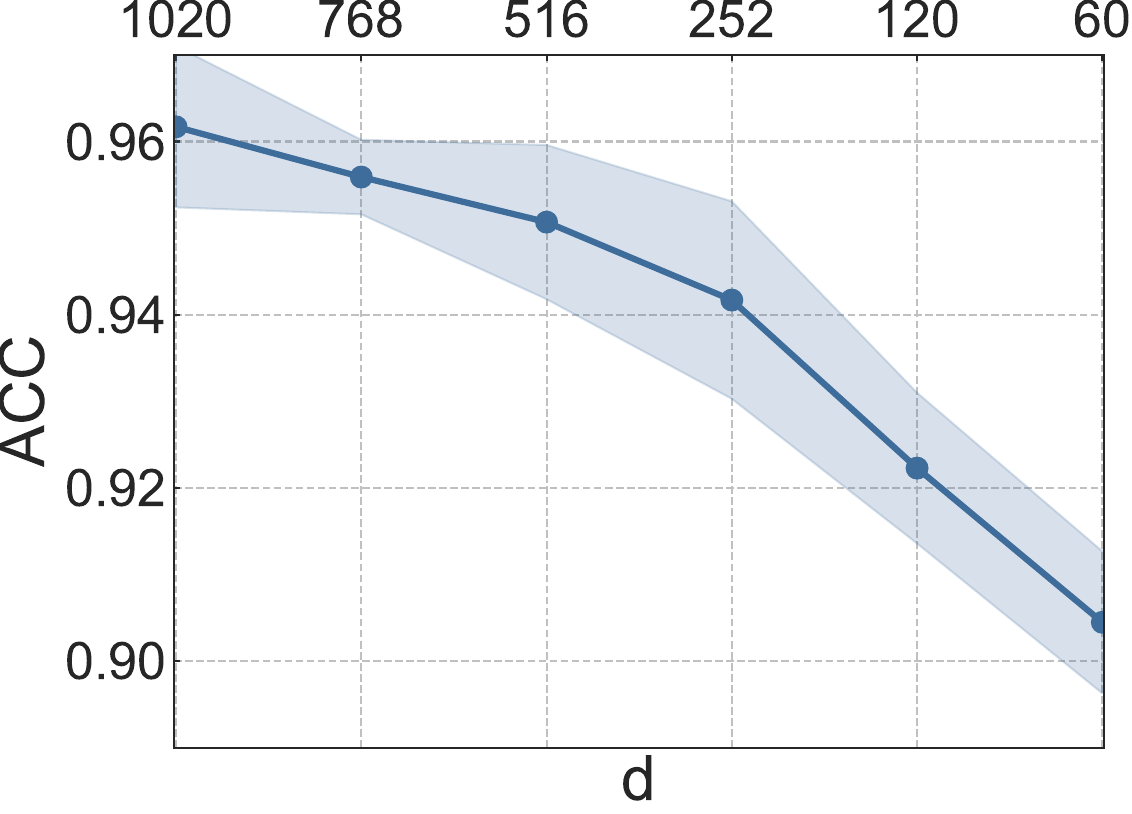}}
\subfloat[Gossipcop]{\includegraphics[width=0.25\textwidth]{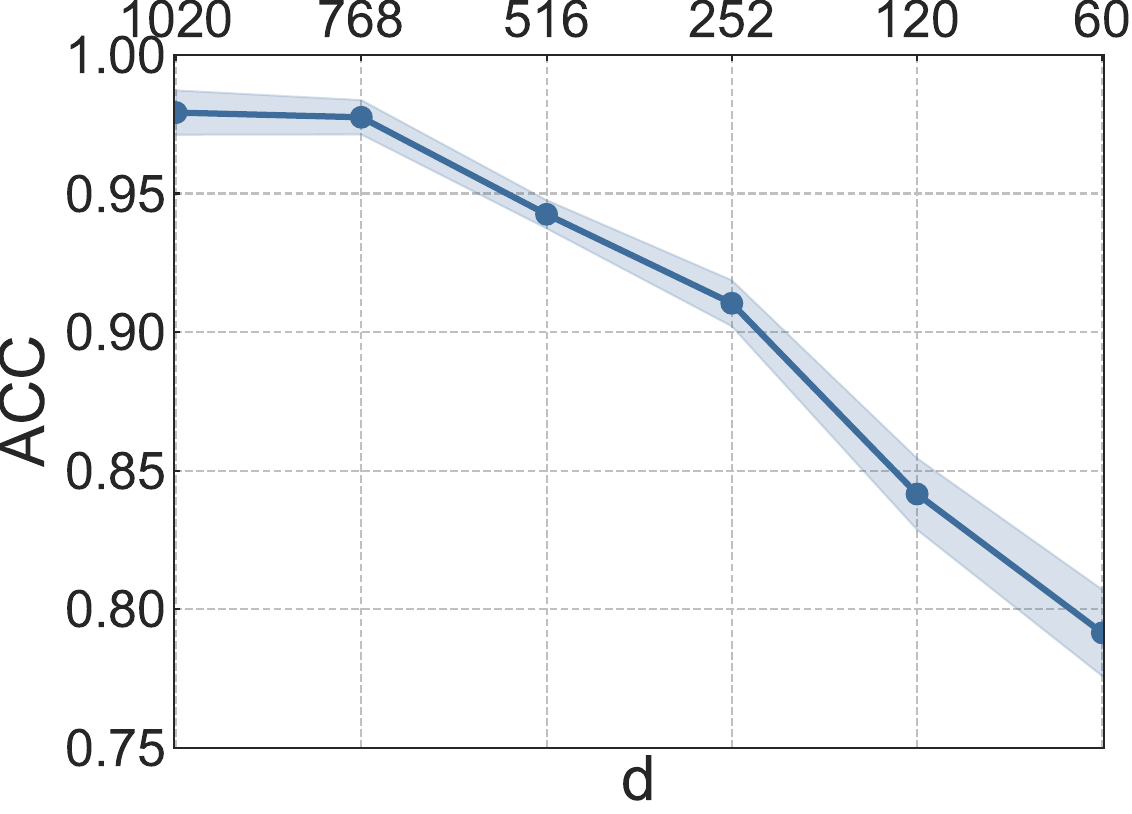}}\\
\subfloat[PolitiFact]{\includegraphics[width=0.25\textwidth]{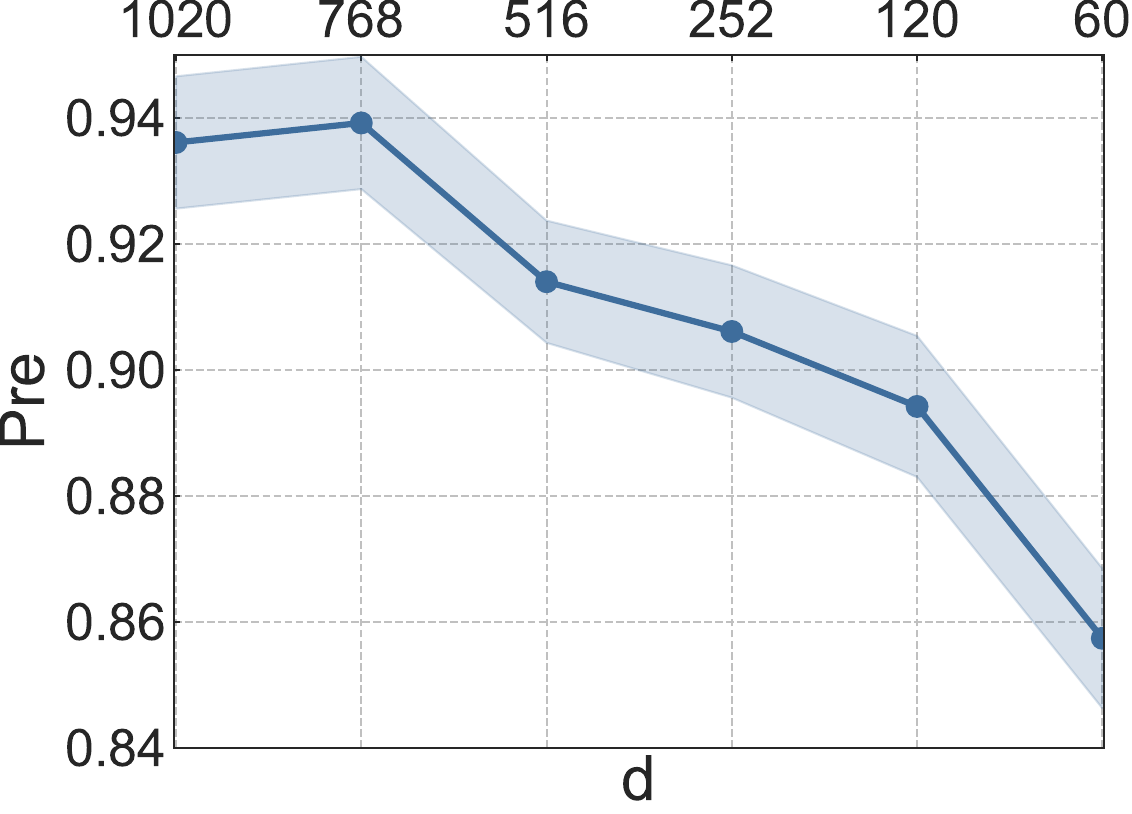}}
\subfloat[ReCOVery]{\includegraphics[width=0.25\textwidth]{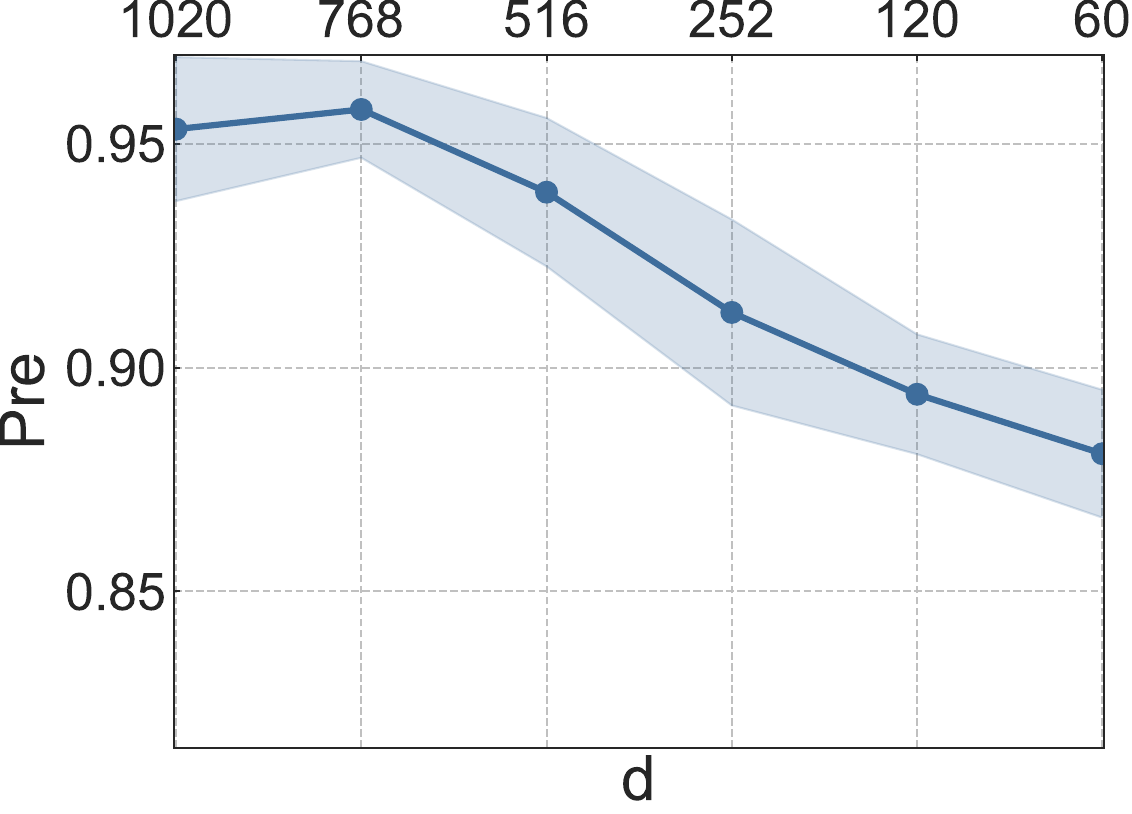}}
\subfloat[MM-COVID]{\includegraphics[width=0.25\textwidth]{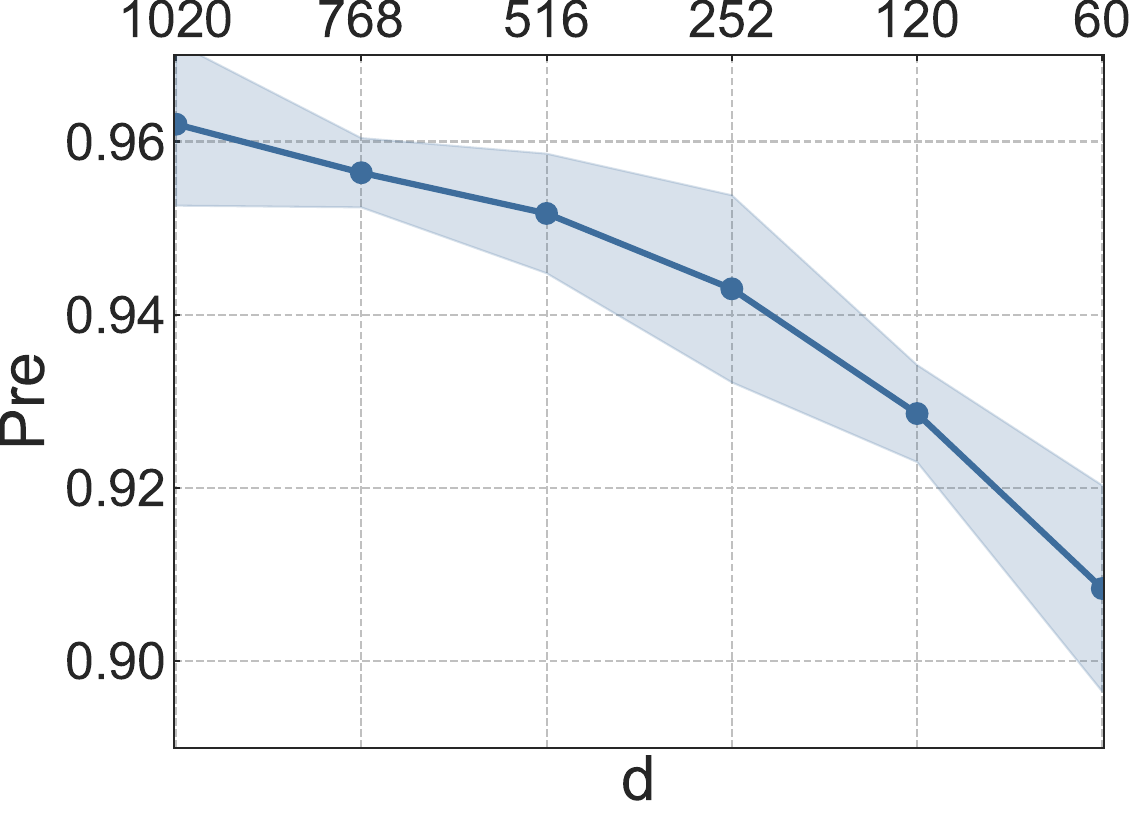}}
\subfloat[Gossipcop]{\includegraphics[width=0.25\textwidth]{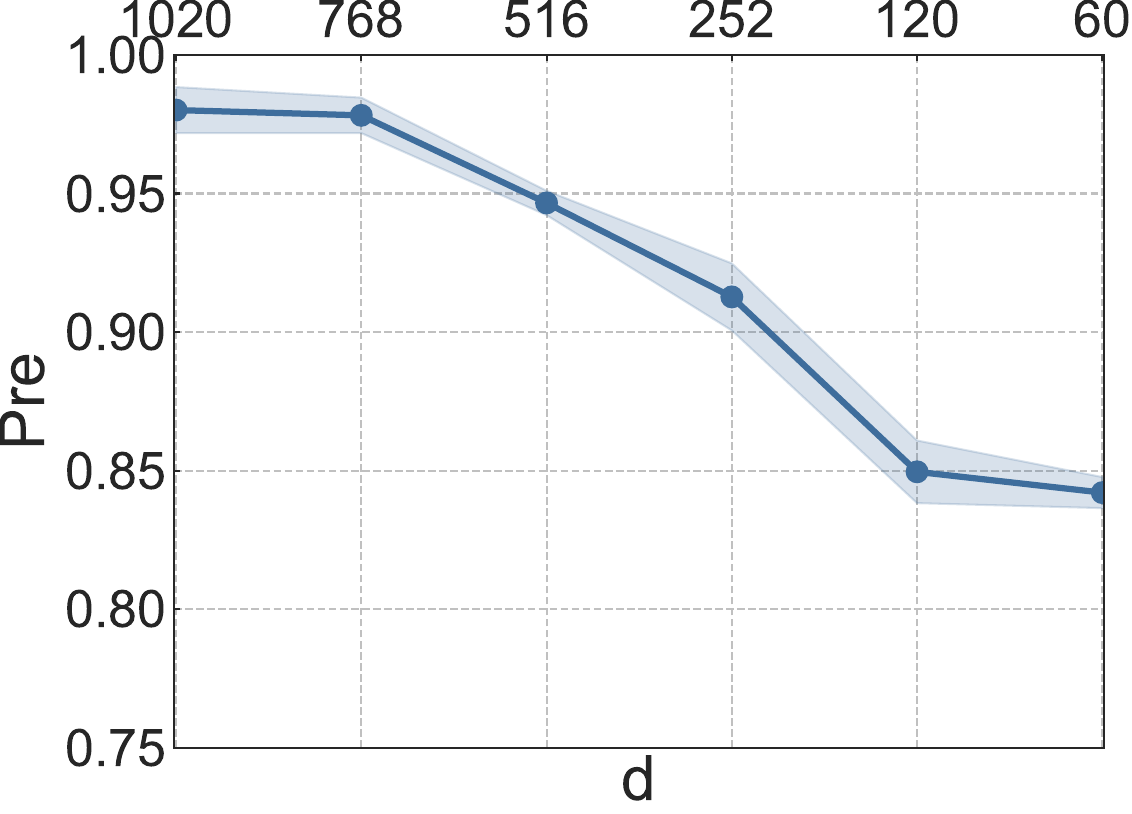}}\\
\subfloat[PolitiFact]{\includegraphics[width=0.25\textwidth]{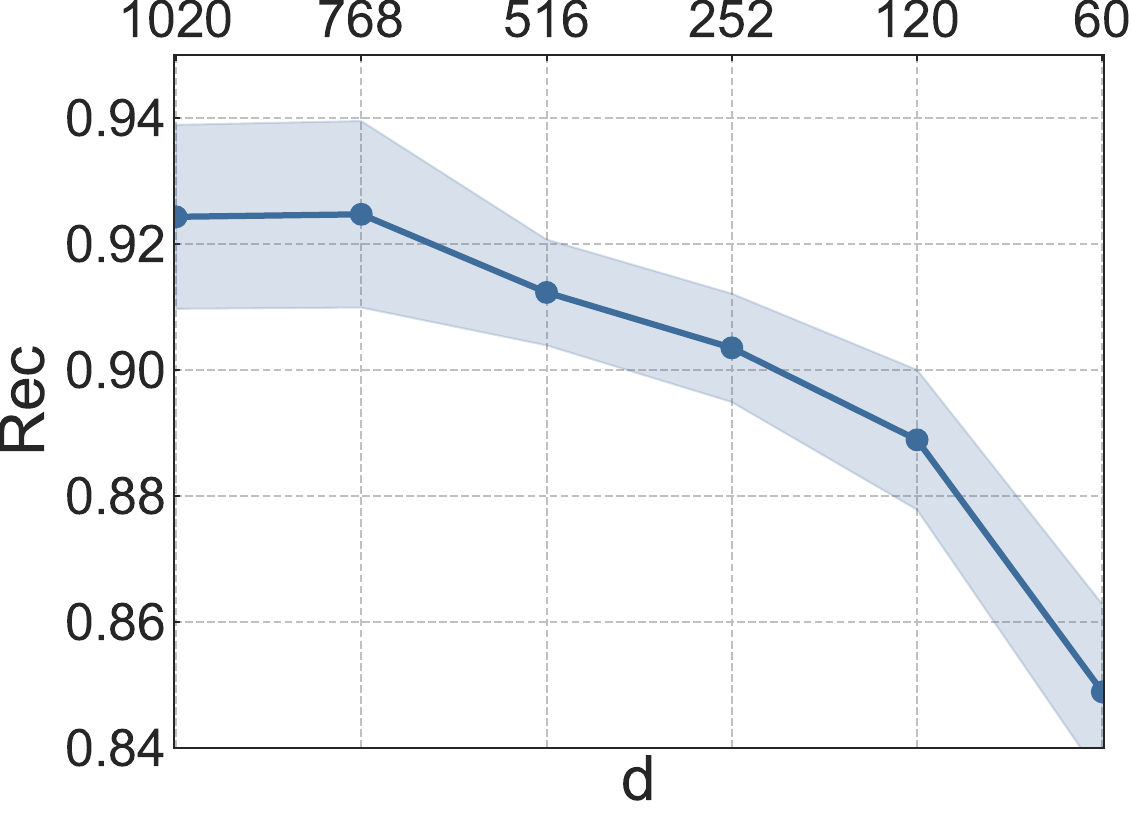}}
\subfloat[ReCOVery]{\includegraphics[width=0.25\textwidth]{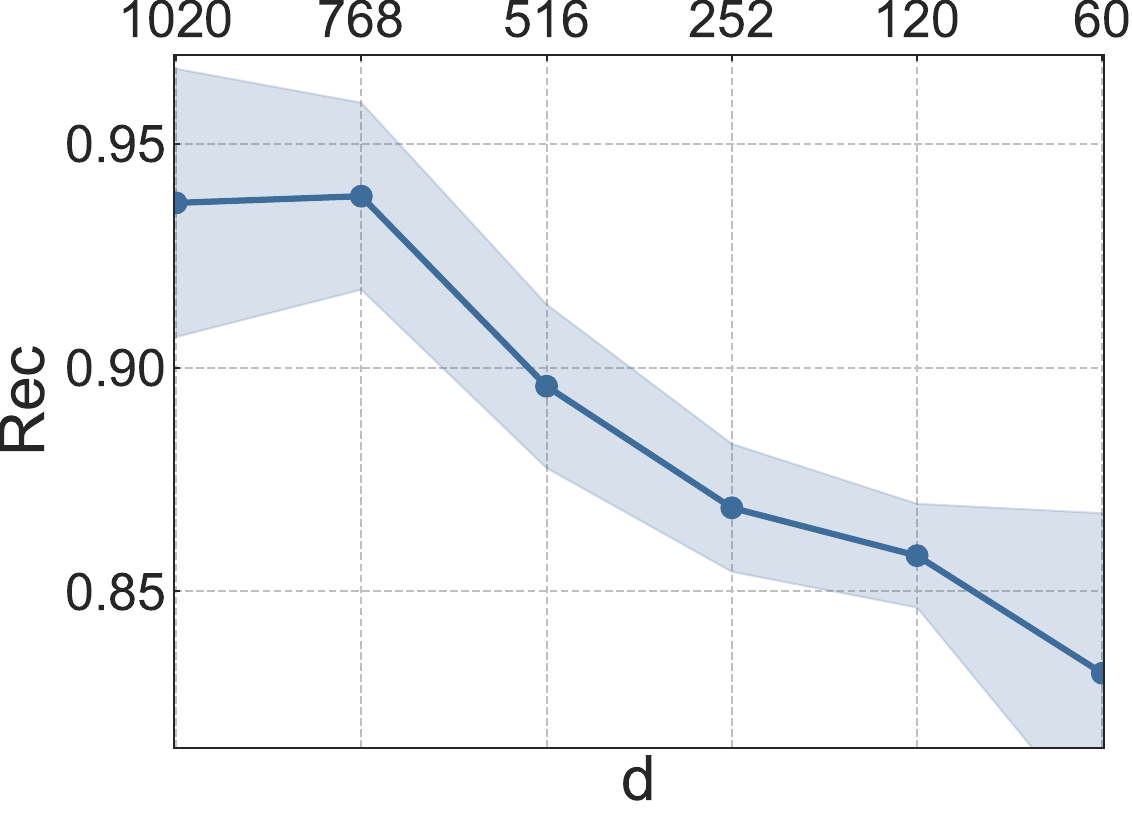}}
\subfloat[MM-COVID]{\includegraphics[width=0.25\textwidth]{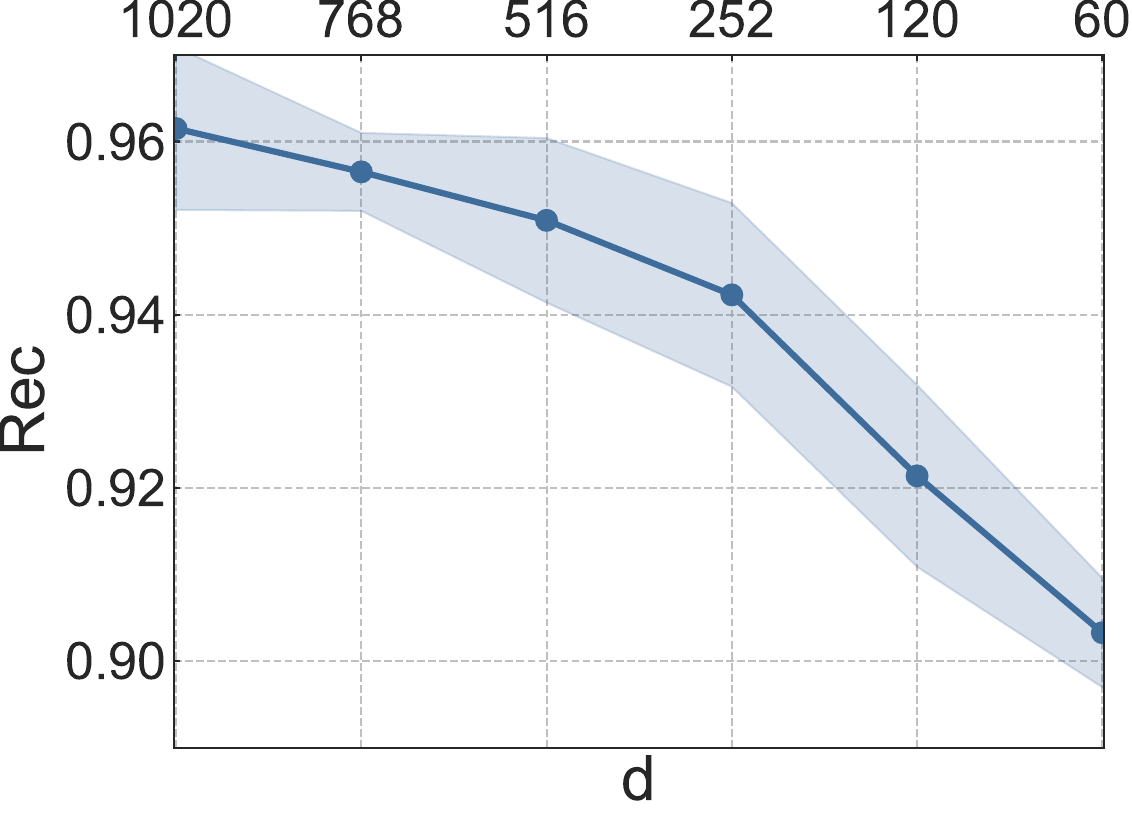}}
\subfloat[Gossipcop]{\includegraphics[width=0.25\textwidth]{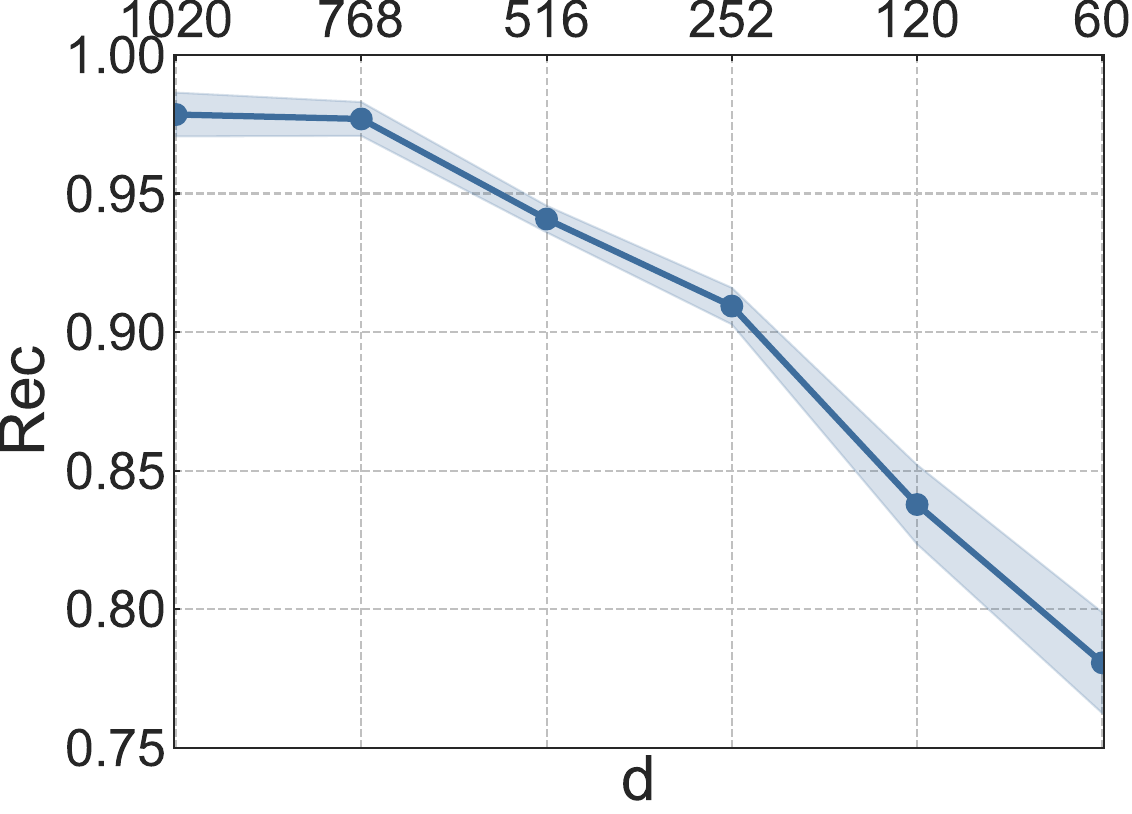}}\\
\subfloat[PolitiFact]{\includegraphics[width=0.25\textwidth]{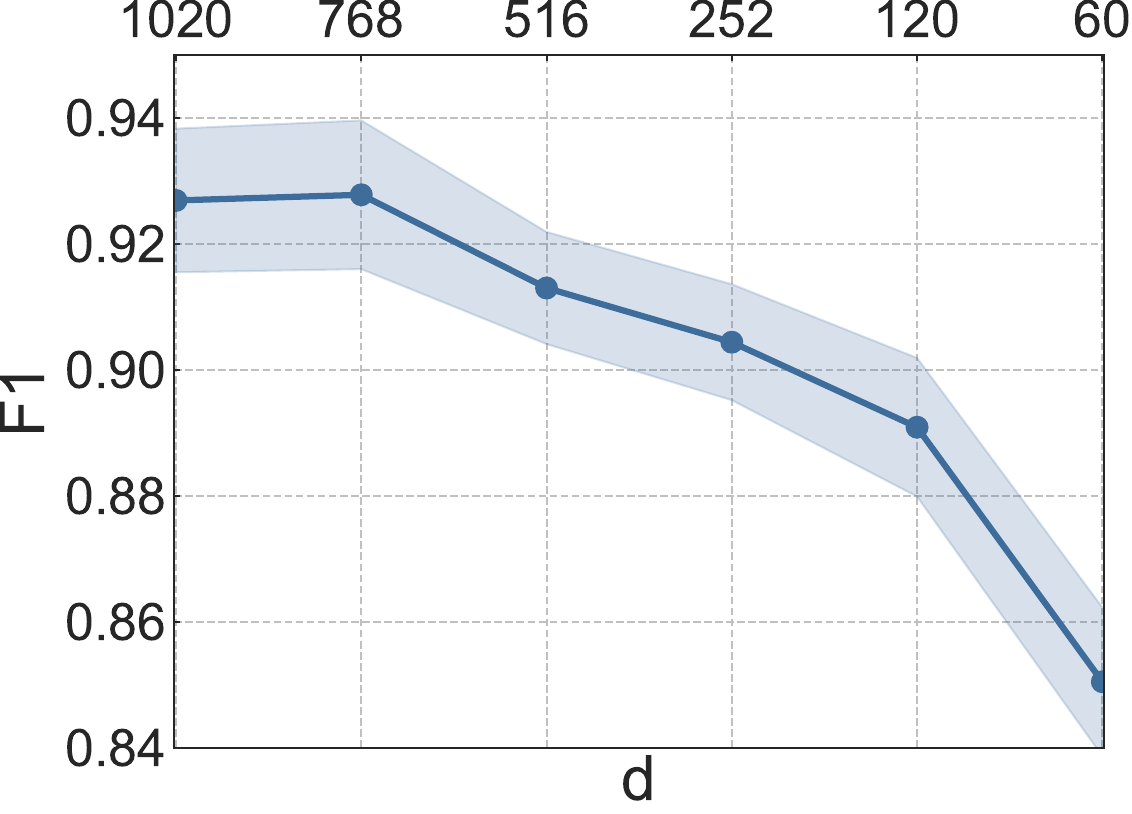}}
\subfloat[ReCOVery]{\includegraphics[width=0.25\textwidth]{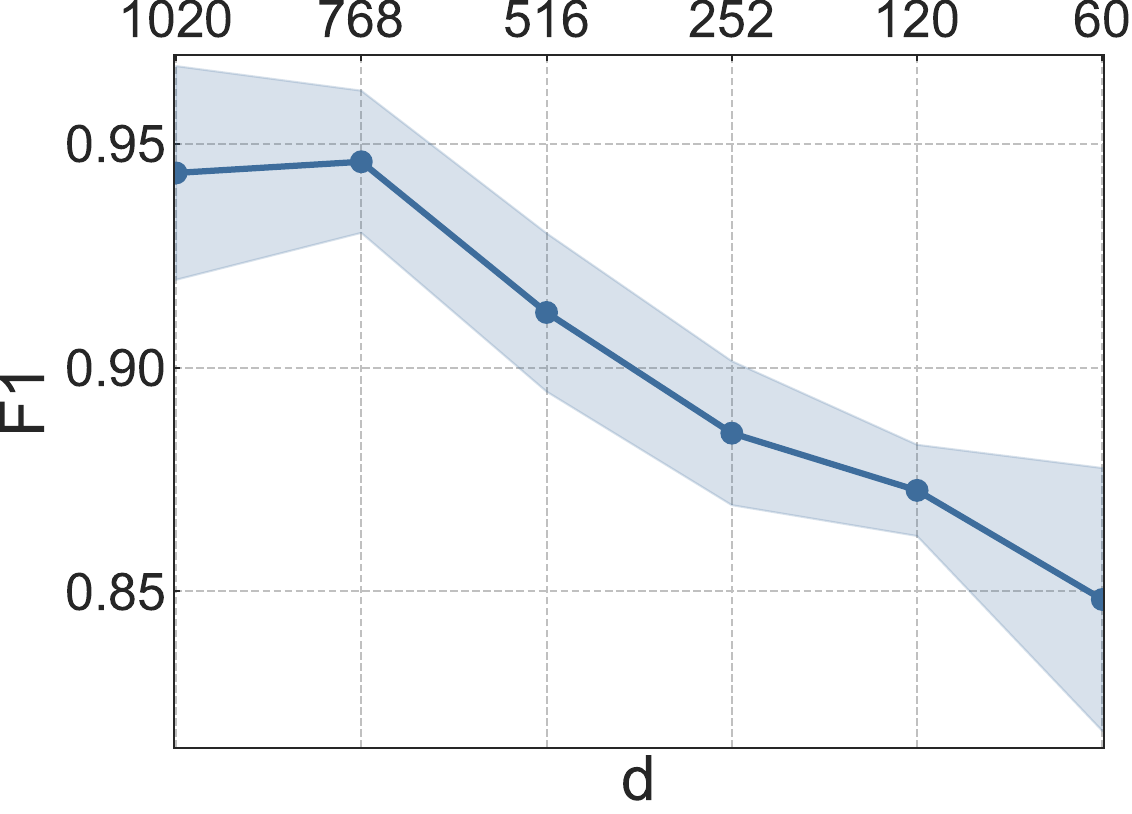}}
\subfloat[MM-COVID]{\includegraphics[width=0.25\textwidth]{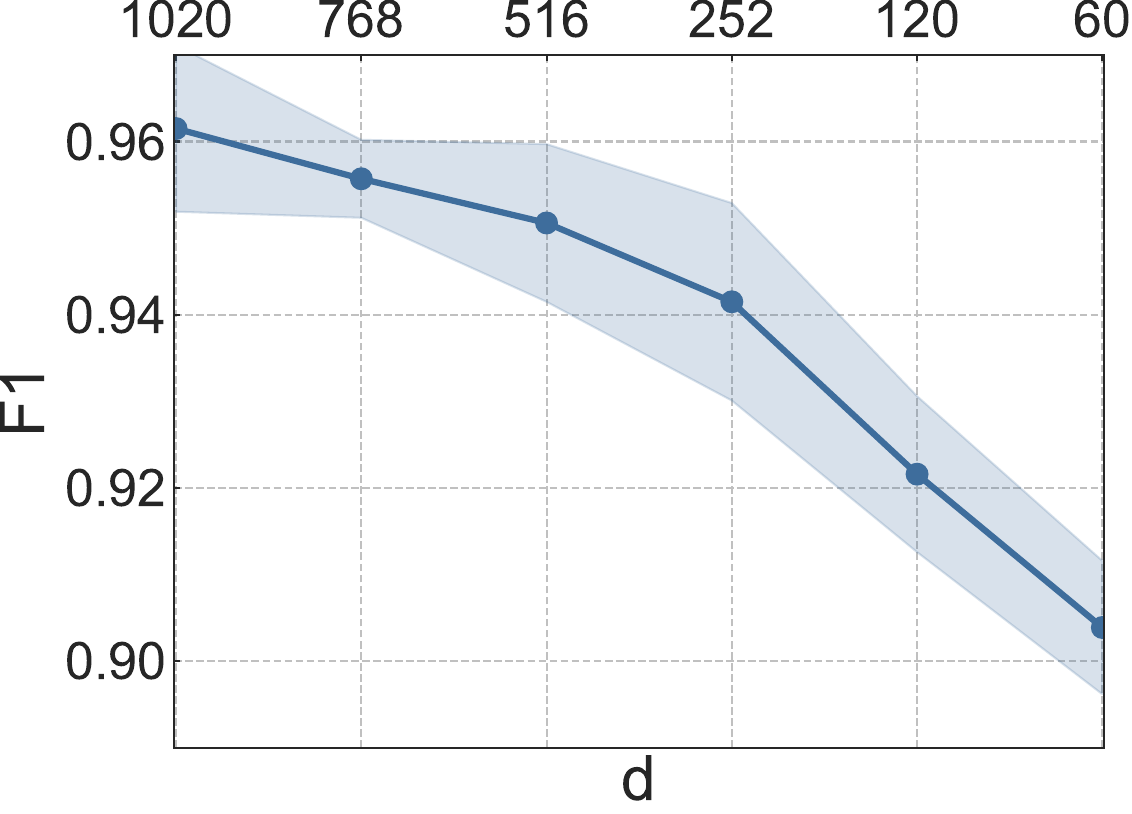}}
\subfloat[Gossipcop]{\includegraphics[width=0.25\textwidth]{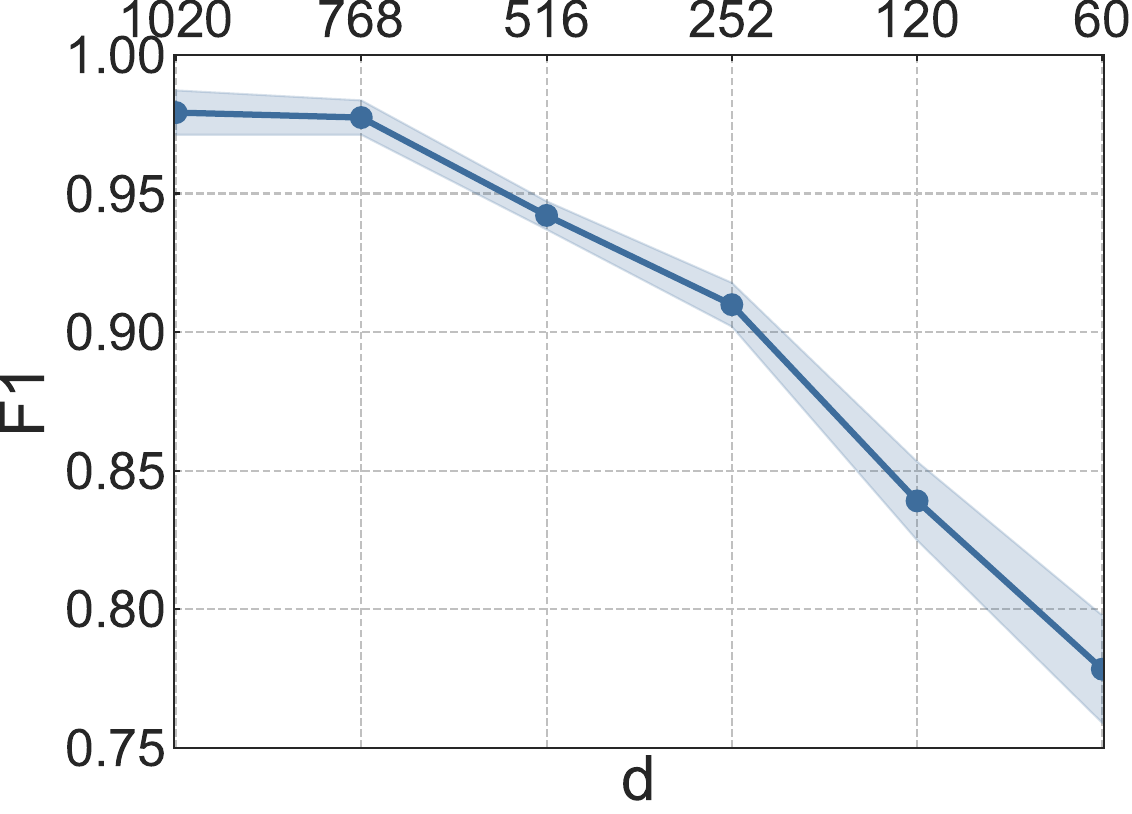}}
\caption{Paramater Analysis.}
\label{fig:parameter}
\end{figure*}

To examine the sensitivity of Hy-DeFake to changes in different parameters, we evaluate the results of different embedding sizes $d$ by the ACC, Pre, Rec, and F1 on four datasets. As shown in Fig. \ref{fig:parameter}, $d$ is set to be the dimensions of news embeddings and user embeddings. It is crucial to note that RoBERTa, which we used to update the embeddings of news, requires that the dimension of embeddings be a multiple of the number of attention heads 12. Moreover, the MI loss in updating the embeddings of news and users in the same dimensions. The results demonstrate that Hy-DeFake performs satisfyingly and stably when the embedding dimensions are 1020, 768, and 516. 

Through the analysis, we set the dimension of 768 as Hy-DeFake's parameter, because our model performs best in most datasets on this dimension, and the default dimension of textual features learned by the language model RoBERTa is 768. Although on the dimension of 1020, our model can also obtain the optimal results in some metrics, the dimension is too high with inefficiency and the result has little difference with the dimension of 768, so we do not set the dimension to 1020. Hy-DeFake performs slightly worse on the 516 dimensions than on the 768 dimensions. When the dimension decreases from 252 to 60, the performance of our model gradually deteriorated. This decline is due to the gradual loss of real latent textual and structural information. Overall, our method maintains satisfactory results across most parameter settings and outperforms baseline methods.

\subsection{Time Efficiency}
To evaluate the execution time of Hy-DeFake, we record its training time on four datasets. Due to the variations in input sizes (\emph{e.g.}, the hypergraph describing users spreading the news in Hy-DeFake is not considered by text-based baselines), comparing the runtime directly with other baseline methods is unfair. TABLE \ref{tab:time} presents the average training time per epoch for Hy-DeFake on each dataset. The results demonstrate that Hy-DeFake achieves efficient and effective training, with superior performance on four datasets. The training time of Gossipcop is the longest, since the large number of users in this dataset. Overall, Hy-DeFake learns the most informative embeddings in an acceptable time and achieves superior and more stable results than baseline methods.  

\begin{table}[htb]
\renewcommand\arraystretch{1.25}
\setlength{\tabcolsep}{1.5mm}
    \centering
    \begin{tabular}{c|cccc}
    \toprule
        \textbf{Datasets} &\textbf{PolitiFact} & \textbf{ReCOVery} & \textbf{MM-COVID} & \textbf{Gossipcop}  \\
    \midrule
         \textbf{AV. Time} & 2.42s & 6.27s & 1.92s & 86.38s\\   
    \bottomrule    
    \end{tabular}
    \caption{Average Training Time on Each Epoch. }
    \label{tab:time}
\end{table}

\subsection{Discussion and Future Work}
\label{sec:discussion}
In this section, we delve deeper into the impact of users in fake news detection. Section of \ref{sec:casestudy} (Case Study and Visualization) reveals distinct differences in both attributes and high-order structures between users who engage with real news versus fake news. The discovery of this phenomenon presents a potential opportunity to identify fake news without text analysis, by leveraging high-order relations among users in social networks, \textit{i.e.}, integrating user attributes with their structural connections. Exploring this opportunity is promising for domain-specific news detection, because investigating this news, such as healthcare news, requires domain knowledge and expertise. However, it is challenging for the public to acquire such knowledge. Thus, in our future work, we will explore whether it is possible to address this issue by capturing high-order structural information without analyzing the textual content of news.

\section{Conclusion}
\label{sec:conclusion}
In this work, our task is to detect fake news in online social networks, and we argue that existing methods face two challenges in doing so. To address these challenges, we explore the high-order correlation between news and users. Firstly we construct an attributed hypergraph to capture the intricate relationships between news and users in online social networks. Subsequently, we propose Hy-DeFake, a hypergraph neural network-based method for detecting fake news. Hy-DeFake not only learns semantic embeddings of news content and credibility embeddings of users, but also incorporates high-order correlations between news and users. By integrating these embeddings, Hy-DeFake can provide informative embeddings for real and fake news classification. Extensive experiments on four real-world datasets demonstrate the superior performance of Hy-DeFake, indicating that it learns distinctive embeddings that contain rich semantic and credibility information as well as high-order correlations. Furthermore, our findings reveal a positive correlation between news authority and user credibility. Users who spread fake news exhibit more intensive interaction compared to those who spread real news, resulting in the formation of a denser community.

In our future work, we aim to further investigate the significance of user credibility in the detection of fake news. Additionally, as discussed in Section \ref{sec:discussion}, we plan to explore the possibility of utilizing high-order relations between news and users to detect fake news without relying on textual analysis in specific domains where distinguishing textual information is challenging due to a lack of professional knowledge.




\bibliographystyle{IEEEtran}
\bibliography{main}

\begin{IEEEbiography}
[{\includegraphics[width=1in,height=1.25in,clip,keepaspectratio]{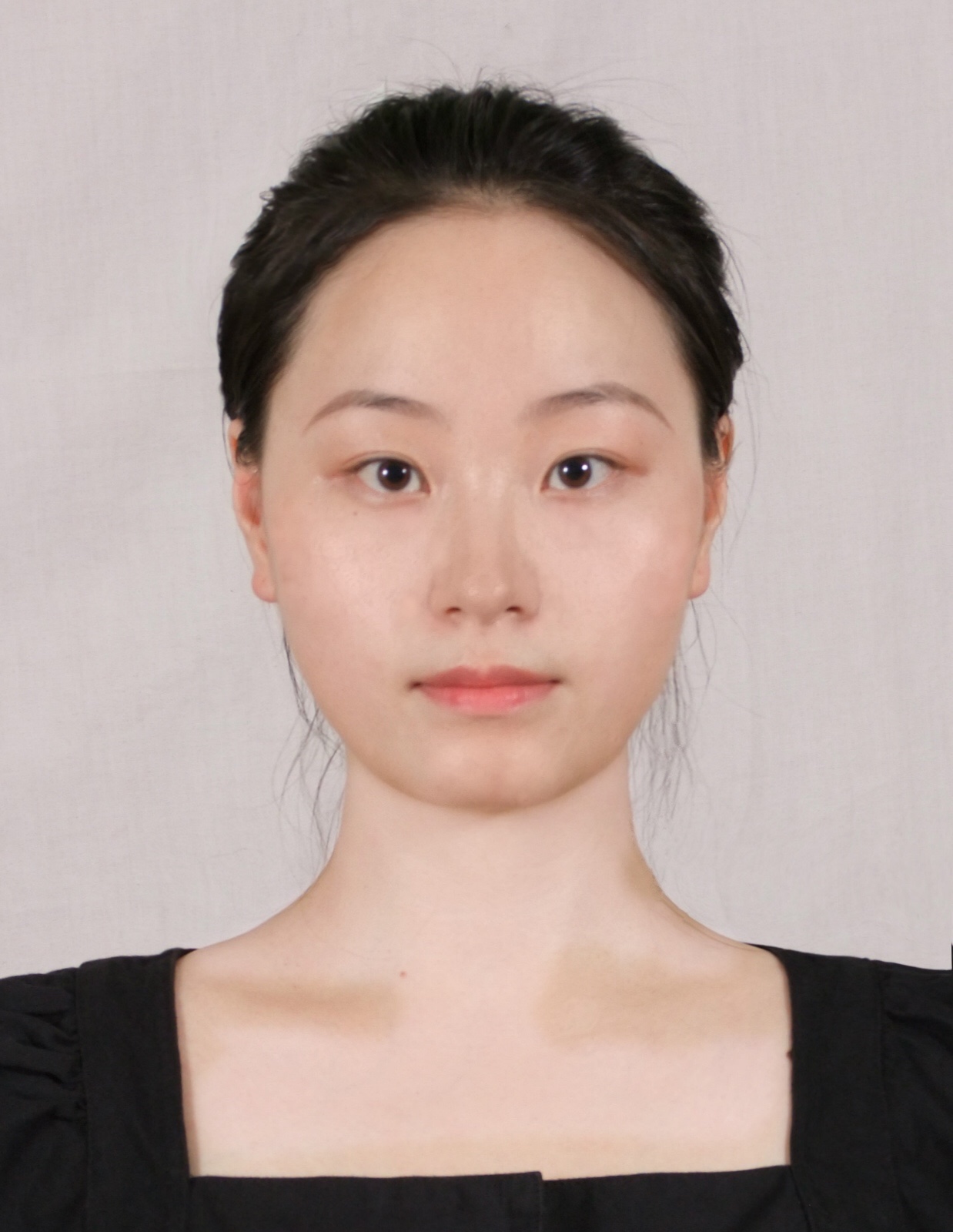}}]{Xing Su} received her M.Eng. degree in computer technology from Lanzhou University, China in 2020. She is currently a Ph.D. candidate in the School of Computing at Macquarie University, Australia. Her current research interests include misinformation detection, community detection, deep learning, and social network analysis.
\end{IEEEbiography}

\begin{IEEEbiography}[{\includegraphics[width=1in,height=1.25in,clip,keepaspectratio]{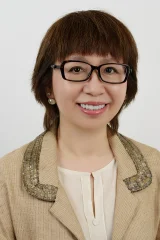}}]{Jian Yang} is a full professor at the School of Computing, Macquarie University. Her main research interests are: graph learning, big data analytics, social networks. Prof. Yang has published more than 200 journal and conference papers in international journals and conferences such as VLDBJ, TKDE, TNNLS, KDD, NeurIPS, IJCAI, VLDB, ICDE, ICDM, etc. She is currently serving as an Executive Committee for the Computing Research and Education Association of Australia.
\end{IEEEbiography}


\begin{IEEEbiography}[{\includegraphics[width=1in,height=1.25in,clip,keepaspectratio]{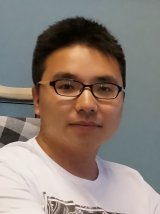}}]{Jia Wu} (M'16) is currently the Research Director for the Centre for Applied Artificial Intelligence and the Director of HDR (Higher Degree Research) in the School of Computing at Macquarie University, Sydney, Australia. 

Dr Wu received his Ph.D. degree in computer science from the University of Technology Sydney, Australia. His current research interests include data mining and machine learning. Since 2009, he has published 100+ refereed journal and conference papers, including TPAMI, TKDE, TKDD, TNNLS, TMM, KDD, ICDM, WWW, and NeurIPS.

Dr Wu has been serving as the Programme Committee Chair/Contest Chair/Publicity Chair/(Senior) Programme Committees for the prestigious data mining and artificial intelligence conferences for over 10 years, such as KDD, ICDM, WSDM, IJCAI, AAAI, WWW, NIPS, CIKM, SDM, etc. His research team was the recipient of the CIKM'22 Best Paper Runner-Up Award, ICDM'21 Best Student Paper Award, SDM'18 Best Paper Award in Data Science Track, IJCNN'17 Best Student Paper Award, and ICDM'14 Best Paper Candidate Award. Dr Wu is the Associate Editor of ACM {Transactions on Knowledge Discovery from Data} (TKDD) and Neural Networks. Dr Wu is a Senior Member of the IEEE.
\end{IEEEbiography}

\begin{IEEEbiography}[{\includegraphics[width=1in,height=1.25in,clip,keepaspectratio]{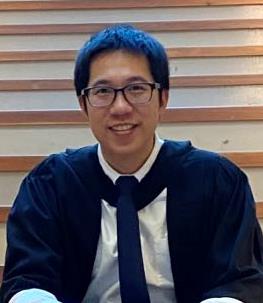}}]{Zitai Qiu}
is currently a Ph.D. student at the School of Computing, Macquarie University, Sydney, Australia. He got his Master's Degree from the University of Queensland, Australia. His research interests mainly include: data mining; deep learning; social event detection and machine learning.
\end{IEEEbiography}

\end{document}